\documentclass[aps,twocolumn,superscriptaddress,reprint]{revtex4-1}
\usepackage{graphicx,amsmath}
\usepackage{subfigure}
\usepackage{epsfig}
\usepackage{url}

\graphicspath{{figures/}}

\begin{document}

\title{Photonic waveguide mode to free-space Gaussian beam extreme mode converter}

\author{Sangsik Kim}
\affiliation{Center for Nanoscale Science and Technology, National Institute of Standards and Technology, Gaithersburg, Maryland 20899, USA}
\affiliation{Maryland Nanocenter, University of Maryland, College Park, MD 20742 USA}
\affiliation{Department of Electrical and Computer Engineering, Texas Tech University, Lubbock, TX 79409, USA.}

\author{Daron A. Westly}
\affiliation{Center for Nanoscale Science and Technology, National Institute of Standards and Technology, Gaithersburg, Maryland 20899, USA}

\author{Brian J. Roxworthy}
\affiliation{Center for Nanoscale Science and Technology, National Institute of Standards and Technology, Gaithersburg, Maryland 20899, USA}

\author{Qing Li}
\affiliation{Center for Nanoscale Science and Technology, National Institute of Standards and Technology, Gaithersburg, Maryland 20899, USA}
\affiliation{Maryland Nanocenter, University of Maryland, College Park, MD 20742 USA}

\author{Alexander Yulaev}
\affiliation{Center for Nanoscale Science and Technology, National Institute of Standards and Technology, Gaithersburg, Maryland 20899, USA}
\affiliation{Maryland Nanocenter, University of Maryland, College Park, MD 20742 USA}

\author{Kartik Srinivasan}
\affiliation{Center for Nanoscale Science and Technology, National Institute of Standards and Technology, Gaithersburg, Maryland 20899, USA}

\author{Vladimir A. Aksyuk}
\email[]{vladimir.aksyuk@nist.gov}
\affiliation{Center for Nanoscale Science and Technology, National Institute of Standards and Technology, Gaithersburg, Maryland 20899, USA}

\begin{abstract}
Integration of photonic chips with atomic, micromechanical, chemical and biological systems can advance science and open many possibilities in chip-scale devices and technology.
Compact photonic structures for direct coupling of light between high-index single-mode waveguides and arbitrary free-space modes spanning hundreds of waves in cross-section would eliminate bulky optical components and enable integration of photonics into many new applications requiring wide beams, structured light and centimeter-scale propagation distances with low diffraction-limited losses. Conventional fiber-coupling approaches do not scale well for accurate, low-loss coupling across the extremely large mode scale mismatch ($\approx10^6$ times in modal area).
Here we present an extreme mode converter that can transform the photonic waveguide mode to the diffraction-limited, free-space Gaussian beam, with a beam waist of about $160~\mu$m.
Using two identical converters, we demonstrate a grating-to-grating coupling that couples the radiating beam back to the chip through a mirror reflection in free-space. Operating at 780~nm for integration with chip-scale atomic vapor cell cavities, our design can be adapted for visible, telecommunication or other wavelengths. Furthermore, other types of beams can be implemented by using the 2-stage expansion approach presented in this paper.
\end{abstract}

\maketitle

\section{Introduction}

Chip-scale photonic devices have advanced fundamental research in atomic physics \cite{kitching2016nist,hummon2018photonic,mehta2016integrated,kohnen2011array}, time/frequency metrology \cite{kippenberg2011microresonator,spencer2017integrated,li2017stably,kim2017dispersion}, and biology \cite{liang2013scalable,fan2011optofluidic,xu2008folded,jokerst2009chip,lin2013trapping}, as well as in industrial applications such as telecommunications \cite{thomson2016roadmap,agrell2016roadmap,jahani2017photonic} and light detection and ranging (LIDAR) \cite{doylend2011two,sun2013large,poulton2017large}.
In many such applications, efficient coupling of nanophotonic circuits with engineered, application-specific free-space optical fields in millimeter scale volumes has opened a broad range of possibilities for chip-scale, highly integrated sensors and systems. For example, the National Institute of Standards and Technology (NIST) is currently implementing chip-scale photonic systems with integrated atomic vapor cavities \cite{kitching2016nist,hummon2018photonic}. Realizing the full potential of such systems requires advances in development of compact, accurate, and efficient optical coupling between sub-micrometer wide photonic waveguide modes and at least 100~$\mu$m wide free-space modes, such as multiple overlapping plane waves, Gaussian, and Bessel beams. The challenge is to decrease the circuit footprint and increase the accuracy of the intensity, phase, and polarization control achieved in this extreme mode conversion, which spans multiple orders of magnitude in mode size.

Grating couplers are the most widely known approach to interfacing a photonic mode and a radiation mode \cite{mekis2011grating,chen2010apodized,ding2013ultrahigh,mehta2017precise,xu2012complementary,song2015polarization,vermeulen2010high,chen2012wideband,halir2010continuously}; the spatial phase modulation of periodic gratings compensates for the momentum mismatch between the photonic and radiation modes. For telecommunication and other applications, typical grating couplers are designed to interface a photonic mode and an optical fiber mode whose mode field diameter (MFD) is about $5~\mu$m to $10~\mu$m (mode area $<10^2~\mu$m$^2$). In such cases, a compact and highly efficient coupler is desirable and various grating designs have been proposed and demonstrated for different polarizations (TE and/or TM) and for different spectral bands (C- and O- bands).
These couplers do not consider in general the phase and intensity profiles of radiation modes but rather are optimized to maximize the power transfer to the fiber, often approximated as a Gaussian mode.
Grating apodization can form the Gaussian mode profiles in the radiation mode and have been used to increase the fiber-to-chip coupling efficiency \cite{chen2010apodized,ding2013ultrahigh} and to focus the radiating beam at a certain distance \cite{mehta2017precise,mehta2016integrated}. The grating apodization requires a careful optimization since varying the duty-cycle will change the effective index; the grating pitch should be adjusted accordingly for the beam collimation.
Also, typical grating couplers are designed in 2D cross-sections and do not consider the varying mode intensity and phase in the lateral direction, that is parallel to the grating lines.
The waveguide is simply tapered out so that the waveguide mode expands to the slab mode with a cone shape.
While producing acceptable losses for fiber coupling, such an approach fails in beam collimation in the lateral direction, which is important for coupling into spatially extended modes. To fully shape the radiation beam and manipulate the radiation direction in 3D free-space, it is essential to design a mode expander that can achieve a specific mode profile and a good beam collimation. Accurate engineering of the desired radiation mode intensity and phase across the 2D plane is required.

In this paper, we present an extreme mode converter that can interface with the photonic mode in a waveguide (modal area of $\approx 300$~nm$~\times~250$~nm) and the Gaussian beam in a free-space (modal area of ~$\approx 160^2~\mu$m$^2$). The mode mismatch between the two modes is about $0.34 \times 10^6$ times (in area).
The extreme mode converter consists of two stages (as in Fig.~\ref{concept}), a waveguide-to-slab mode expander followed by an apodized grating. First, a $160~\mu$m wide, collimated (one-dimensional) Gaussian slab mode is created, and then a large apodized grating with straight lines is used to couple it to free space. Separating the two stages and producing a collimated slab mode with a flat wavefront in the first stage effectively makes the second stage apodization problem two-dimensional, and therefore analytically and numerically tractable. Optimizing the spatially varying period and duty cycle of the grating achieves the desired Gaussian intensity and flat wavefront in the orthogonal direction at the $2^{\rm nd}$ stage. Analytical and numerical methods are combined for the design and optimization. We experimentally demonstrate the chip-to-beam conversion and characterize the mode intensity profile and the wavefront of the generated beams by capturing the real and Fourier images, respectively. We also study the out-coupling angles (polar and azimuthal) of the beam and achieve grating-to-grating coupling, which couples the radiating beam back to the chip by using two extreme mode converters and placing a flat mirror a few millimeters above and parallel to the chip .

\section{Modeling: two-stage extreme mode converter}
\begin{figure}[!t]
\hspace{-6mm}
\includegraphics[width=0.51\textwidth]{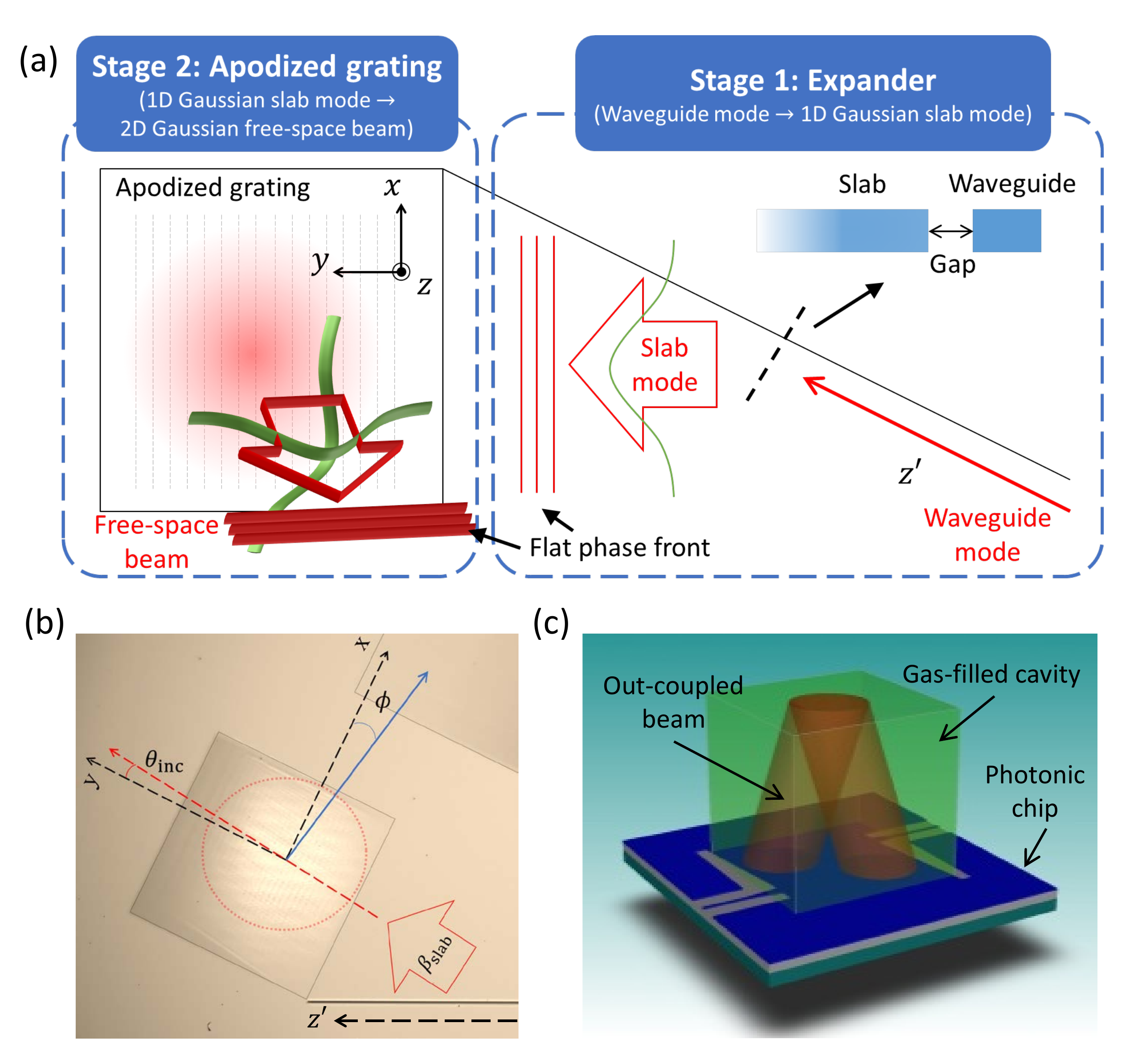}
\caption{Extreme mode converter.
(a) Schematic of extreme mode converter with two step conversions (stage 1: photonic waveguide mode to 1D-Gaussian slab mode, stage 2: 1D-Gaussian slab mode to 2D-Gaussian beam in free-space).
(b) Microscope image of the fabricated extreme mode converter with a coordinate system ($xy$-axes, incident angle $\theta_{\rm inc}$, and azimuthal angle $\phi$).
The propagation direction of the waveguide mode is defined as $z'$.
(c) Concept figure of a photonic chip with two extreme mode converters coupling light in and out for optical interrogation (red) of a gas-filled cavity volume (green).
A mirror placed on top of the gas-filled cavity is used to reflect the beam radiated from one grating into the second grating.
}
\label{concept}
\end{figure}
The extreme mode converter consists of two stages: stage 1 is a mode expander that converts the photonic waveguide mode into the slab mode, and stage 2 is the apodized grating that couples out the slab mode into the free-space mode. Figure~\ref{concept}(a) gives the schematic of these conversions.
To achieve the extreme mode conversion, the expander in stage 1 is designed to attain a wide 1D-Gaussian intensity profile with a flat wavefront in the slab, and the grating in stage 2 is apodized by varying both grating period and duty-cycle to realize the 2D-Gaussian beam with a large beam waist ($w_0>100~\mu$m).
Specifically, the period is varied such that the phase of the out-coupling beam is designed to be flat, to achieve a high-quality beam collimation.
Figure~\ref{concept}(b) shows a microscope image for the fabricated converter and with definitions of the coordinates and angles. The incident angle $\theta_{\rm inc}$ is defined for the angle between the slab mode $\beta_{\rm slab}$ and the grating vector $\frac{2\pi}{\Lambda}$ ($y$ axis).
The azimuthal angle $\phi$ is also defined with respect to the grating lines ($x$ axis).
Figure~\ref{concept}(c) illustrates an application, where a chip with two extreme mode converters is integrated to optically interrogate a gas-filled cavity in a compact system. To experimentally demonstrate this concept, we place a mirror above the chip surface, at the plane of the mode overlap from the two converters, to characterize the grating-mirror-grating coupling. Our devices are designed to operate at the free-space wavelength of $\lambda_0=780$~nm, and silicon nitride (Si$_3$N$_4$) is chosen for the photonics guiding material to minimize the losses at this wavelength \cite{kitching2016nist}.

\subsection{Stage 1 (expander): waveguide mode to 1D-Gaussian slab mode conversion}
\begin{figure}[!t]
\begin{center}
\hspace{-6mm}
\includegraphics[width=0.51\textwidth]{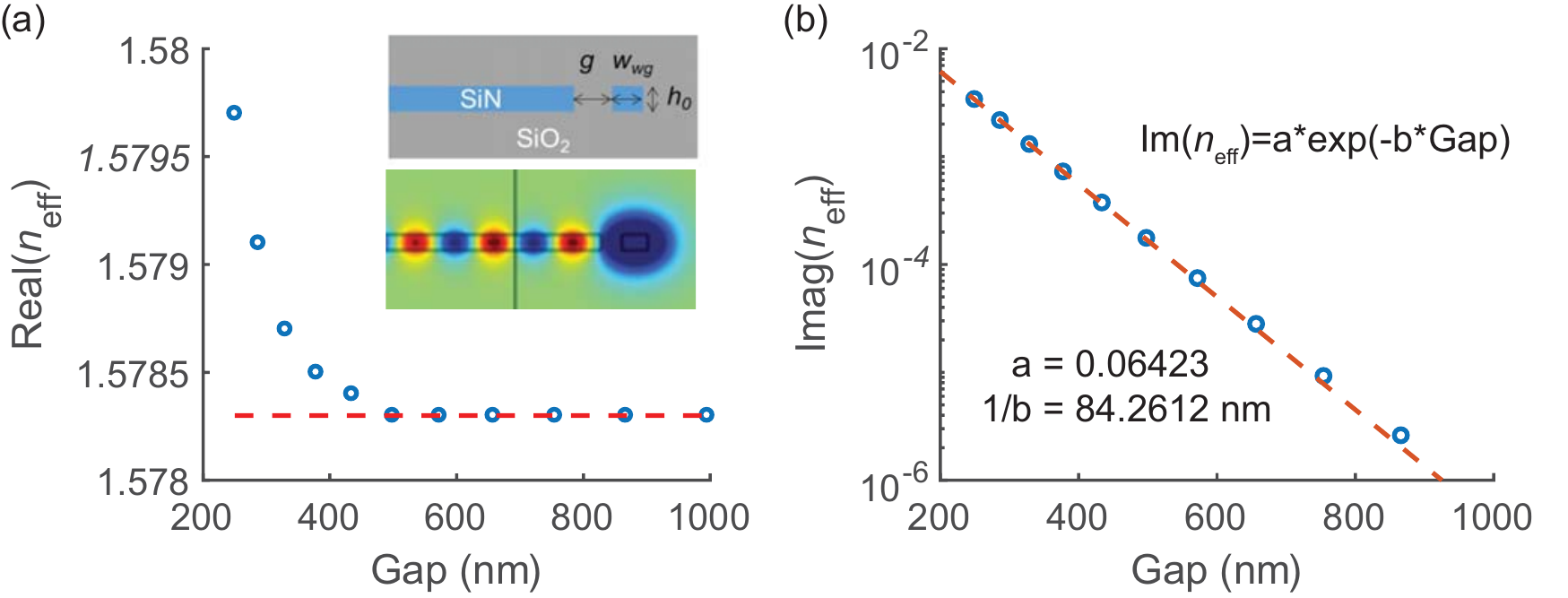}
\end{center}
\caption{Stage 1: photonic waveguide mode to 1D-Gaussian slab mode conversion by evanescent coupling. Numerically calculated effective refractive index ($n_{\rm eff}$) of the photonic waveguide mode as a function of the gap size $g$ between the waveguide and the slab: (a) Real part of $n_{\rm eff}$ (dashed orange line: limiting value for $g>500$~nm, inset: cross-sections of schematic and FEM domain with computed TE mode profile) and (b) Imaginary part of $n_{\rm eff}$ (dashed orange line: fitting curve, text: fitting parameters). The height $h_0$ and the width $w_{\rm wg}$ of the waveguide are 250~nm and 300~nm, respectively.
}
\label{stage1}
\end{figure}

Stage 1 is a mode expander that converts the photonic waveguide mode to the slab mode with a flat wavefront and a 1D-Gaussian lateral power density distribution in $x$ direction (the expander part in Fig.~\ref{concept}(a)). The basic principle for the mode expander is evanescent coupling. The coupling strength between the waveguide and the slab depends on the gap size $g$ between them, and we can design the gap profile $g(z')$ to form a Gaussian intensity distribution in the slab (along the waveguide direction of the wave propagation $z'$). First, to quantify the coupling strength between the waveguide and the slab, we used a finite element method (FEM) to numerically evaluate the complex effective refractive indices ($n_{\rm eff}$) as a function of $g$. We use a commercial FEM solver in the frequency domain to calculate the waveguide cross-sectional mode profile and its effective index. The power from the waveguide couples to the slab and is radiated in-plane, which results in an imaginary component of the index, accounting for the mode power decay along the expander. To model this within a finite geometrical domain, and to avoid complications of using perfectly matched layers (PML) within an eigenmode calculation, we choose to introduce optical losses to the slab material instead. The portion of the slab closest to the waveguide is modeled as a perfect lossless dielectric, and further away from the waveguide the slab material optical loss is increased adiabatically. This ensures no reflection of the slab mode back toward the waveguide from either within the slab or from the domain boundary (Figures~\ref{stage1}(a) inset, Supplementary movie 1).

Figures~\ref{stage1}(a) and \ref{stage1}(b) are the real and imaginary parts of the simulated $n_{\rm eff}$, respectively. Insets in Fig.~\ref{stage1}(a) show the schematic of the simulation domain and the mode profile of the fundamental transverse-electric (TE$_0$) mode (See Supplementary for the transverse magnetic mode data, movie 2). All the devices in this paper are designed based on the TE$_0$ mode.
The thickness of the Si$_3$N$_4$ layer (waveguide and slab) $h_0$ is set to 250~nm and the waveguide width $w_{\rm wg}$ is set to 300~nm. All elements made of Si$_3$N$_4$ are clad with the SiO$_2$ (upper: $>1~\mu$m, lower: $\approx 2.9~\mu$m).

The $\operatorname{Re}(n_{\rm eff})$ corresponds to the propagation constant $\beta=\operatorname{Re}(n_{\rm eff})\beta_0$ of the waveguide mode ($\beta_0=2\pi/\lambda_0$ is the free-space wavevector)
and determines the tilt-angle $\theta_{\rm tilt}$ of the slab mode direction of propagation relative to the waveguide. For our $h_0$, the effective index of the 1D TE slab mode is calculated to be 1.79 and the tilt angle between the waveguide and slab modes can be estimated by $\theta_{\rm tilt}=\cos^{-1}(\operatorname{Re}(n_{\rm eff})/1.79)$. Note that in Fig.~\ref{stage1}(a), the $\operatorname{Re}(n_{\rm eff})$ approaches 1.578 for $g>500$~nm. This indicates that the evanescent coupled slab mode would have the same tilt angle $\theta_{\rm tilt}=\cos^{-1}(1.578/1.79)=28.18^{\circ}$ for gap sizes larger than 500~nm. In other words, the evanescent-coupled waves in the slab will be collimated if the $g(z')>500$~nm.
In Fig.~\ref{stage1}(b), $\operatorname{Im}(n_{\rm eff})$ corresponds to the power loss of the evanescent coupling and decreases approximately exponentially as the gap size increases.
Using this data, and assuming adiabatic variation of the gap profile $g(z')$, we can design $g(z')$ for the desired power distribution along $z'$. The phase at the slab boundary is the same as the phase in the waveguide, and increases linearly along $z'$ for a constant $\operatorname{Re}(n_{\rm eff})$. For $g(z')>500$~nm, the variation in gap size does not shift the $\theta_{\rm tilt}$ and only affects the power distribution. Smaller gaps can be employed, but variation in $\operatorname{Re}(n_{\rm eff})$ may have to be compensated. Curving the slab boundary and the waveguide appropriately can be used to achieve the desired wavefront for the slab mode. In our design the slab edge is straight, creating a flat wavefront for the collimated slab mode.

We now need to achieve the correct intensity profile to obtain the  Gaussian slab mode, and we have used the following procedures to design the $g(z')$. The optical power in the waveguide, $P(z')$ can be written as
\begin{equation}
\frac{dP(z')}{dz'} = -P(z') \alpha(z') ,
\label{eq1:Pwg}
\end{equation}
where $\alpha(z')$ is the loss coefficient that can be written as $\alpha(z')=\frac{4\pi}{\lambda_0}\operatorname{Im}(n_{\rm eff})$. We can set initial power in the waveguide $P(-\infty)=1$.
We want the power density in the slab $\frac{dP_{\rm s}(z')}{dz'}$ to form a Gaussian mode with a beam waist of $w$, and it can be represented as the following:
\begin{equation}
\frac{dP_{\rm s}(z')}{dz'} = C\exp\left(\frac{-2z'^2}{w^2}\right).
\label{eq2:Ps}
\end{equation}
The coefficient $C=\frac{1}{w}\sqrt{\frac{2}{\pi}}$ can be obtained by setting the total power as 1 and integrating the $\int_{-\infty}^{\infty}\frac{dP_{\rm s}(z')}{dz'}dz'=1$. For the energy conservation, the total power in the waveguide and the slab should be equal to 1, $i.e., P(z')+P_s(z')=1$, and the loss in the waveguide should be equal to the coupling power of the slab at that segment, $i.e., dP_{\rm s}(z')=-dP(z')$. Rewriting these two conditions, we have the following equations:
\begin{equation}
P(z')=1-\int^{z'}_{-\infty}\frac{1}{w}\sqrt{\frac{2}{\pi}}\exp\left(\frac{-2\zeta^2}{w^2}\right) d\zeta,
\label{eq3}
\end{equation}
\begin{equation}
\frac{1}{w}\sqrt{\frac{2}{\pi}}\exp\left(\frac{-2{z'}^2}{w^2}\right) = P(z') \frac{4\pi}{\lambda_0} \operatorname{Im}(n_{\rm eff}).	
\label{eq4}
\end{equation}
Solving these two equations and using the relation of $\operatorname{Im}(n_{\rm eff})=a\exp\left(-bg(z')\right)$, where $a$ and $b$ are the fitting coefficients from Fig.~\ref{stage1}(b), we can derive the gap profile $g(z')$ as the following:
\begin{equation}
g(z')=\frac{1}{b} \ln \left\{ \frac{1}{a} \frac{\lambda_0}{\sqrt{2}\pi^{3/2}w}
\frac{ \exp(-2{z'}^2/w^2) }{1-\operatorname{erf}( \sqrt{2}z'/w )} \right\}.
\label{eq5}
\end{equation}
Note that the beam waist $w$ is of the Gaussian distribution along the waveguide direction. The actual beam waist $w_0$ of the resulting 1D-Gaussian slab mode, normal to its direction of propagation in the slab, is obtained by $w_0=w\sin(\theta_{\rm tilt})$. In our design, we have set the beam waist to be $w_0=100\sqrt{2}~\mu$m.

\begin{figure*}[!ht]
\begin{center}
\includegraphics[width=0.82\textwidth]{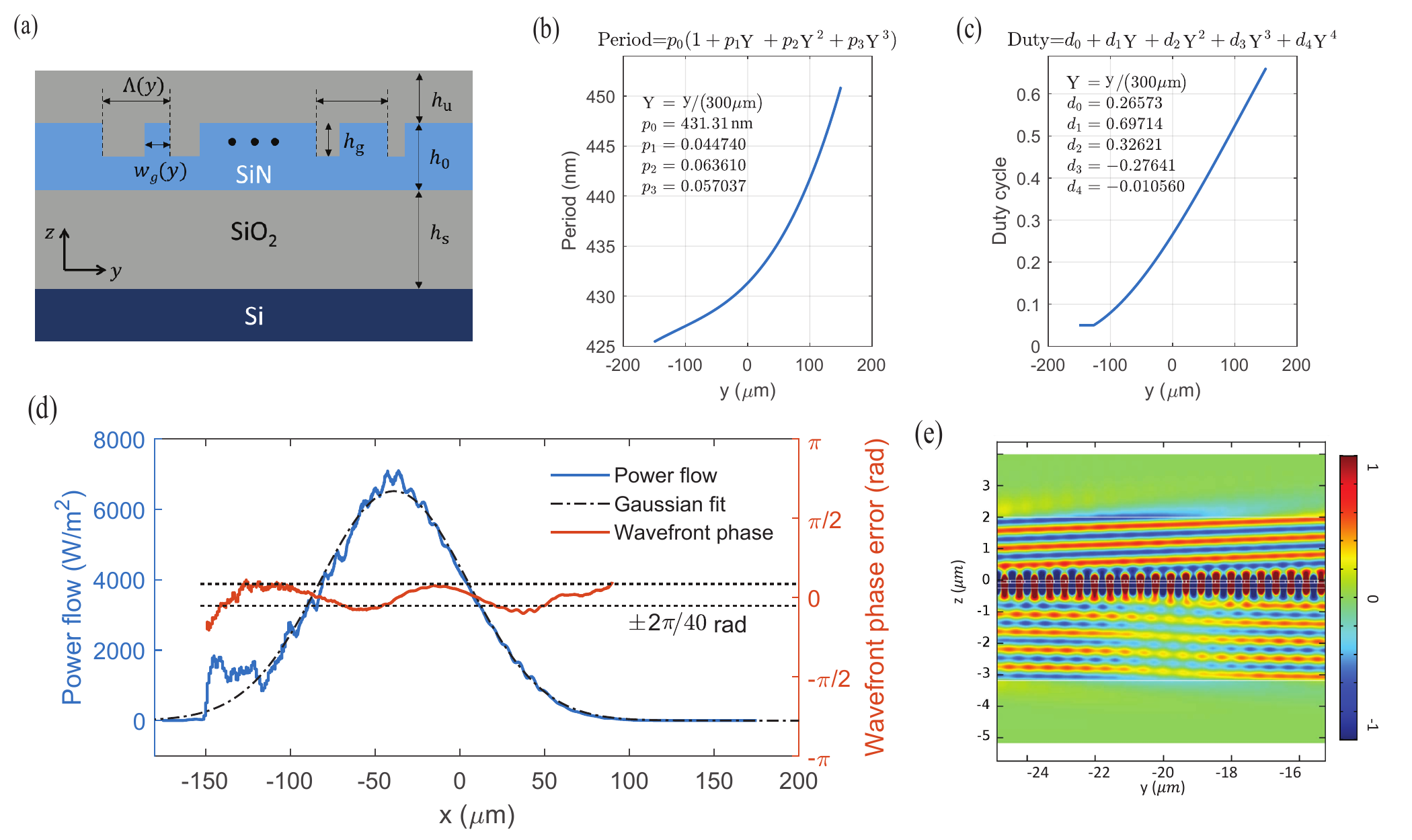}
\end{center}
\caption{Stage 2: 1D-Gaussian slab mode to 2D-Gaussian beam conversion.
(a) Schematic of the apodized grating with geometric parameters: $h_0=250$~nm, $h_{\rm g}=85$~nm, $h_{\rm s}=2.9~\mu$m, and $h_{\rm u}=2.8~\mu$m.
Grating period $\Lambda(y)$ and grating width $w_g(y)$ are apodized.
Numerically optimized grating (b) period $\Lambda(y)$ and (c) duty-cycle $w_g(y)/\Lambda(y)$
(Insets: optimized polynomial coefficients).
(d) FEM results of the out-coupled beam: power flow (blue), Gaussian fit (black dash-dot line), and wavefront phase error (orange).
Dashed lines indicate the upper and lower bounds of the phase error, which are within $\pm2\pi/40$.
(e) Electric field profile ($E_x$) within a portion of the FEM simulation domain.
}
\label{stage2}
\end{figure*}

\subsection{Stage 2: 1D Gaussian slab-mode to 2D Gaussian beam conversion}
Stage 2 is an optimized apodized grating, with spatially-varying duty cycle and period, that out-couples the 1D Gaussian slab mode into the 2D free-space Gaussian. The grating lines are straight and parallel. The slab mode is collimated, so that the phase is invariant along the grating lines, while the intensity is varying only gradually. Therefore, to create the Stage 2 out-coupler, it is sufficient to solve a 2D problem, with translational invariance along the grating lines, creating a collimated Gaussian profile in the plane normal to the grating lines (in $y$ direction).

For a specific case of a collimated Gaussian output, further simplifications could have been applied leveraging the slow variation of the intensity and phase across the grating. However, the 2D TE scattering problem from the slab mode into the free space can be quickly and accurately solved for the $\approx 300~\mu$m grating using a commercial finite element frequency domain solver. This makes it possible to apply a more general numerical optimization technique to solve the inverse problem of finding a grating design that optimizes the coupling between the input slab mode and any arbitrary prescribed free-space mode ($i.e.$ doing the $``$inverse design$"$).

In addition to the electromagnetic (EM) fields, in the same solver, we have introduced the spatially-dependent deformation vector field $(u,v)$, discretized on the mesh. This allows us to continuously and smoothly deform the model, together with the mesh, avoiding digital noise and calculation overhead associated with discrete re-building and re-meshing of the grating model. This is the key to efficient numerical optimization, because not only the EM fields can be numerically computed for a particular deformation, but the gradients of any EM-dependent cost function with respect to all geometrical parameters can be computed cost-effectively as well. This allows the application of efficient gradient-based nonlinear optimization methods, such as sparse nonlinear optimizer (SNOPT), already implemented in the optimization add-on package \cite{ComsolOpt}.

We have defined the grating geometry using 11 scalar variable parameters. Two parameters define the grating etch depth and the thickness of the SiO$_2$ layer separating the Si$_3$N$_4$ grating from the Si wafer. The spatially-dependent grating duty cycle is described by a 4$^{\rm th}$ order polynomial function of the location $y$ (Fig.~\ref{stage2}(c)), while the grating period is given by a 3$^{\rm rd}$ order polynomial of $y$ (Fig.~\ref{stage2}(b)). Polynomial coefficients represent the other 9 variables defining the geometry. We have additionally constrained the duty cycle from decreasing below 0.05 ($\approx20$~nm), to account for the nanofabrication limits on producing extremely narrow grating lines. Qualitatively, the spatially-varying duty cycle, together with the grating depth, control the strength of the local optical coupling between the slab mode and free space. The spatially-varying period ensures that the Gaussian wavefront is planar by compensating for the duty-cycle-dependent effective index of the slab mode. The varying oxide depth ensures that the reflection from the Si wafer constructively interferes, maximizing the optical power in the upward direction.

While it is possible to apply the prescribed deformations only to the model geometric boundaries and obtain a smooth mesh deformation by solving for the $``$numerically induced$"$ deformation at the internal mesh points \cite{ComsolOpt}, we have further reduced the computation complexity by explicitly defining all deformations everywhere in the model as linear interpolations between the prescribed vertical and horizontal displacements of the grating boundaries. I.e., the deformation fields everywhere are explicit functions of the 11 deformation variables, creating desired model deformations described by the two polynomials and two thickness parameters.

We have maximized the modulus square of the S parameter (scattering matrix element) describing the optical coupling between the slab mode input port and a Gaussian mode output port defined on the horizontal domain boundary in free-space above the grating (backed up by a perfectly matched layer to eliminate reflection). The vertical domain size is chosen to be large enough for the evanescent fields from the grating to decay before reaching the boundary. The 2D (cylindrical) Gaussian mode waist center was constrained to the grating surface, but allowed to shift in the $y$ direction along the grating. The Gaussian waist center location, the width and the angle are used as variable optimization parameters. Allowing the Gaussian waist width to vary prevents the optimization algorithm from being stuck in the local optima associated with the angularly-narrow, spatially-wide Gaussian port matching a sharp side-lobe of the extended grating out-coupling pattern. We emphasize that within the gradient-based optimization using deformed geometry, adding extra variables does not drastically increase the computation time. The Gaussian width was forced to $w_0=100~\mu$m by adding a term to the optimization cost function maximizing the waist, while constraining the waist from above to $100~\mu$m. [$w_0=100~\mu$m is the prescribed field waist $E\sim\exp(-x^2/w_0^2)$; the corresponding Full-Width-Half-Maximum (FWHM) is $=w_0\sqrt{2\ln 2}=117~\mu$m]

Figure~\ref{stage2}(a) shows the schematic of the apodized grating in $yz$-axis with spatially-varying period $\Lambda(y)$ and grating width $w_g(y)$. The duty-cycle is defined as $w_g(y)/\Lambda(y)$.
The thicknesses of each layer are $h_0=250$~nm, $h_{\rm g}=85$~nm, $h_{\rm s}=2.9~\mu$m, and $h_{\rm u}=2.8~\mu$m.
Figures~\ref{stage2}(b) and \ref{stage2}(c) show the optimization results for the grating period and duty-cycle, respectively. The insets in each figure show the optimized polynomial coefficients.
Figure~\ref{stage2}(d) shows the numerically simulated out-coupling power flow (blue), its Gaussian fit (black dash-dot line), and the wavefront phase error (orange), when the geometric parameters of Figs.~\ref{stage2}(b) and \ref{stage2}(c) are used.
Figure~\ref{stage2}(e) is the zoomed-in view of the simulated electric field profile ($E_x$).
The resulting optimal out-coupling angle of the Gaussian is 2.2$^{\circ}$ in free-space, and the wavefront error is less than $2\pi/20$ rad over the beam. The power distribution fits well with the Gaussian fit, however the beam FWHM is $\approx103~\mu$m, which is $\approx15$ \% lower than the desired outcome. The port width is forced to the FWHM$~= 117~\mu$m and there is a small coupling penalty associated with this width mismatch. We speculate that the optimization algorithm may be balancing this mismatch loss with additional losses (increased wavefront error and loss into the substrate) associated with extending the Gaussian, or alternatively, that the parameters of the optimization algorithm are not set up perfectly, resulting in a small residual error. The calculated Gaussian port coupling is 68 \%; the power flowing down into the substrate is 26 \%; the slab mode reflection and transmission are negligible. The remaining power accounts for the mode mismatch with the Gaussian port.

Due to the constructive/destructive interference with the downward-outcoupled light back-reflected up from the Si wafer surface, the numerical simulation shows periodically varying upward coupling efficiency from $\approx 67$ \%  for the optimum oxide thickness down to $\approx 35$ \% for $\approx 150$~nm thicker or thinner oxide. As the destructive interference decreases the out-coupled intensity, the optical power in the slab mode propagates further on the grating in the $y$ direction, resulting in a significant widening of the out-coupled Gaussian beam. Experimentally, it is likely that we have had some oxide thickness variation between different runs, which contributes to the observed Gaussian beam width variation in the $y$ direction.

\section{Experiment 1: Gaussian beam characterization}

To demonstrate the extreme mode converter designed in Section~2, we have fabricated and tested the devices.
The fabrication starts with a 100~mm diameter silicon wafer on which thermal oxide is grown. The design target thickness is $2.9~\mu$m, however experimentally this may have varied by as much as 100~nm or more between different fabrication runs, which contributes to run-to-run variability in device performance. In the following step, an approximately $250$~nm thick nominally stoichiometric silicon nitride (Si$_3$N$_4$) layer is deposited by low pressure chemical vapor deposition (LPCVD) and patterned twice. First, electron beam lithography is used to define $\approx300$~nm wide waveguides, inverse-tapers for coupling to the optical fiber and the Stage~1 expander comprising a variable gap between the waveguide and a slab. The nitride is patterned by a reactive ion etch all the way through the layer. In the second electron beam patterning step, the apodized gratings of Stage~2 are defined, and the grating groves are etched nominally 85~nm deep into the nitride layer. A $2.8~\mu$m thick layer of silicon dioxide is deposited by plasma-enhanced chemical vapor deposition (PECVD) after which the wafers are diced and the edges of the chips are polished to expose the ends of the inverse-tapered waveguide fiber couplers.
For the device characterization, monochromatic laser light ($\lambda_0\approx780$~nm) is coupled from an optical fiber to the waveguide mode (TE$_0$) through a tapered fiber-to-chip edge coupler. The extreme mode converter transforms the TE$_0$ mode to the free-space Gaussian beam, and we have characterized the mode intensity profile and the wavefront of the out-coupled Gaussian beam by measuring the microscope images in the real and Fourier spaces, respectively.

\begin{figure}[!hbt]
\begin{center}
\includegraphics[width=0.48\textwidth]{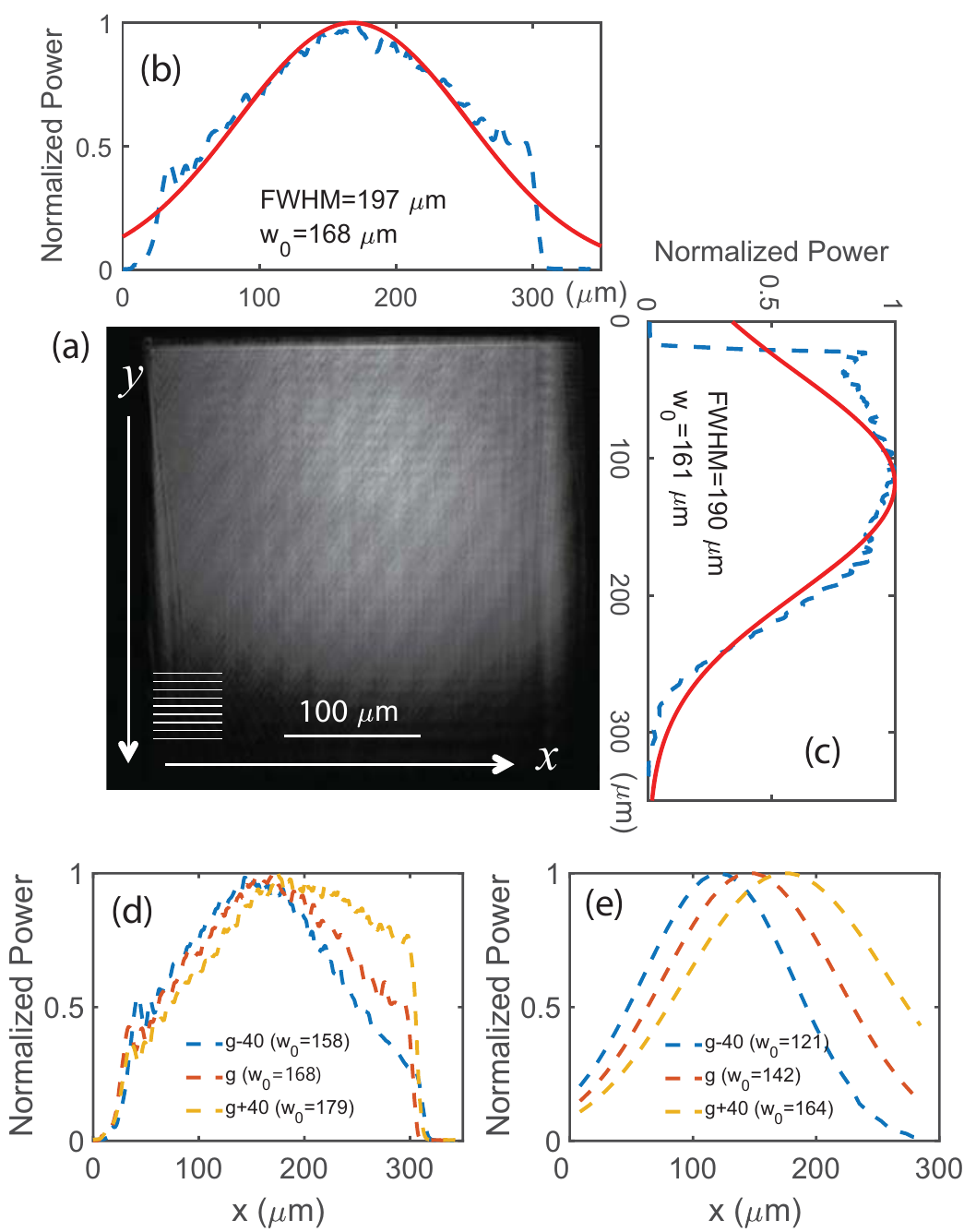}
\end{center}
\caption{Gaussian mode profile on a grating. (a) Microscope image of a converted Gaussian beam on a 300~$\mu$m x 300~$\mu$m grating. The grating lines are parallel to the $x$-axis (as schematically indicated). The scale is calibrated based on the known physical size of the grating. (b) and (c) are the projected images of (a), showing the Gaussian mode profiles along the $x$ and $y$ axes, respectively (blue dashed lines: data, red solid lines: fitting curve). The Full-Width-Half-Maximum (FWHM) and the beam waist $w_0$ are shown in each figure. (d) Measured mode profiles (projected on $x$-axis) for different gap sizes which have constant $\pm$40~nm variations on the gap profile $g(z')$ (blue: $g$-40~nm, orange: $g$, and yellow: $g$+40~nm). (e) Numerically calculated (ODE) mode profiles that are similar to (d). The uncertainties in the characterized beam waist are about $\pm1~\mu$m as determined by the Gaussian fit.
}
\label{profile}
\end{figure}
\subsection{Free-space mode intensity profile characterization}
To characterize the mode profiles of the converted Gaussian beam, we have captured the microscope images of the beam on the grating. Figure~\ref{profile}(a) shows the microscope image of the converted Gaussian beam. The inset scheme shows the direction of the gratings, $i.e.$, the grating lines are parallel to the $x$-axis and perpendicular to the $y$-axis. The slab mode is incident from the top of the image. Figures~\ref{profile}(b) and \ref{profile}(c) are the normalized powers of the beam that are integrated along $y$ and $x$, and projected to the $x$ and $y$ axes, respectively. The dashed blue lines are the data and the red lines are the reference Gaussian curves. The FWHM and the beam waist $w_0$ of each projection are also shown in each figure. Notice that the power distributions of the beam fit well with the Gaussian curves within the grating area and that the beam waist in both $x$ and $y$ axes is $w_0\approx160~\mu$m, which is reasonably close to the design target values ($w_0=100\sqrt{2}=141.4~\mu$m). The fabrication imperfections and the index differences between the modeling and the real materials may have caused these errors. In Fig.~\ref{profile}(c), the Gaussian shape is cut at the beginning part of the gratings; this is due to the minimum feature size limit of the grating (20~nm) in the fabrication, which is similar to the FEM result in Fig.~\ref{stage2}(b).

Uniformly increasing or decreasing $g(z')$ shifts the center position of the slab mode and the Gaussian mode in the grating (along the $x$ axis). It also affects the $w_0$ and FWHM in $x$. Figures~\ref{profile}(d) and \ref{profile}(e) are the measured and numerically calculated intensity mode profiles that are projected to the $x$ axis. For the numerical calculation, we have solved Eq.~\ref{eq1:Pwg} using the ordinary differential equation (ODE) solver with the gap profile $g(z')$ of Eq.~\ref{eq5}. The orange line is the original design with the $g(z')$, and the blue and yellow lines are cases which have the gap uniformly decreased or increased by 40~nm, respectively, $i.e.$, $g(z')-40$~nm and $g(z')+40$~nm. Notice that in both the experimental and numerical cases, the narrower gap size shifts the center position to the left (closer to the beginning of the evanescent coupler). A narrower gap increases the coupling, increasing the slab mode intensity at the beginning of the coupler, and therefore less light remains in the waveguide to couple out toward the end, decreasing the slab mode intensity there. This is opposite for a wider gap. Furthermore, the smaller gap size gives the narrower $w_0$, and again, this is opposite for the larger gap size. In the numerical results, $g(z')+40$~nm gives $w_0=164.0~\mu$m, which is similar to the beam waist from the real experiment ($w_0\approx168~\mu$m$~\pm1~\mu$m); such gap variation is a possible reason for the  $\approx 15~\%$ width difference between the designed $w_0=141.4~\mu$m and the experiments.

\subsection{Wavefront characterization}
\begin{figure*}[!ht]
\begin{center}
\includegraphics[width=1.0\textwidth]{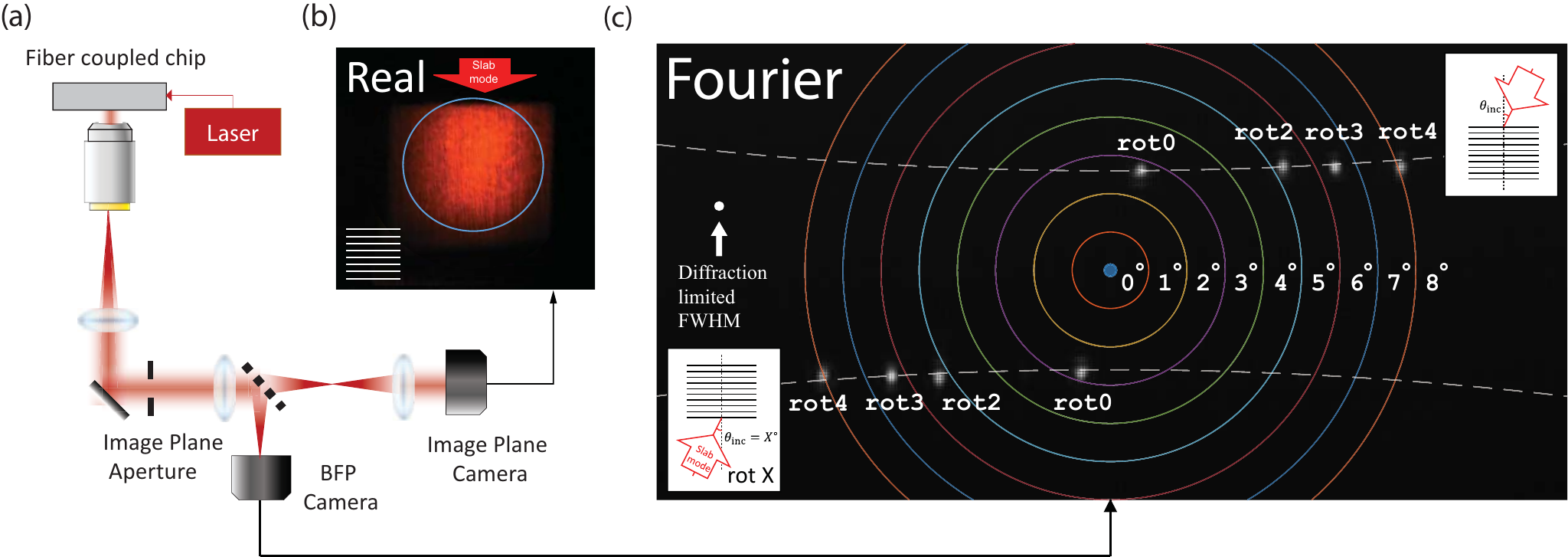}
\end{center}
\caption{Back-focal-plane (BFP) measurement. (a) BFP measurement setup to characterize the beam profile in both angular (Fourier) space and real-space. (b) Real-space image of the converted Gaussian beam (blue line: image plane aperture with a diameter of 250~$\mu$m). (c) BFP composite image of the Gaussian beams for different grating rotational angles relative to the incident slab mode (rot0: $90^{\circ}-\theta_{\rm tilt}=61.82^{\circ}$ rotation from the expander waveguide, ${\rm rot2}={\rm rot0}+2^{\circ}, {\rm rot3}={\rm rot0}+ 3^{\circ}$, and ${\rm rot4}={\rm rot0}+4^{\circ}$). BFP image X (Y) axes are oriented along (normal to) the grating lines in (b). Two identical devices rotated 180$^{\circ}$ are measured to establish origin (surface-normal). The white dashed lines are theoretical out-coupling angles using the Eq.~\ref{eq7} and Eq.~\ref{eq8}. A white dot indicates the ideal diffraction-limited beam's full width at half maximum (FWHM).}
\label{wavefront1}
\end{figure*}
To check the beam collimation of the converted Gaussian mode, we have measured its far field intensity as a function of angle (Fourier space) by capturing the back focal-plane (BFP) images. Figure~\ref{wavefront1}(a) shows the measurement setup for simultaneous Fourier and real-space imaging. Monochromatic laser light at $\lambda_0\approx780$~nm was coupled to the chip through glued, edge-coupled single-mode fiber inputs. Two charge coupled device (CCD) cameras were placed at the real and BFP image planes to capture the real- and Fourier-space images, respectively. A movable variable-diameter circular aperture was placed in the image plane of the microscope, before the beam splitter, allowing for full or spatially-selective evaluation of the far field light, $i.e.$ coming from the whole grating or any specific part of the grating selected by the aperture.
Figure~\ref{wavefront1}(b) shows a real image at the grating, and the blue line indicates the outline of the aperture with a diameter of $\approx250~\mu$m. Figure~\ref{wavefront1}(c) shows the angular (Fourier) space images at the BFP; the colored lines represent the grid (one degree/line) of the polar angle $\theta$. The gray spots are the actual beam images from each device. Notice that all the spot sizes are quite small and near diffraction-limited; otherwise, the spots would spread broadly over wider angles.
The white dot is a guide for the eye with diameter indicating the ideal diffraction-limited beam's FWHM.
Note that for each device set, we have two identical mode converters with opposite orientation ($i.e.$, mode converter~1 is rotated $180^{\circ}$  relative to the mode converter 2); thus, in Fig.~\ref{wavefront1}(c), the upper four gratings are in opposite orientations to the lower four gratings. The rot0 refers to a device whose grating lines are nominally orthogonal to the incident slab mode, which is designed to have a tilt angle $\theta_{\rm tilt}=28.18^{\circ}$ relative to the waveguide. The rot2, rot3, and rot4 are devices with an additional 2$^{\circ}$, 3$^{\circ}$, and 4$^{\circ}$ rotation of the grating relative to the incident slab mode, respectively. We have rotated the gratings to engineer the Gaussian beam polar angle and direction. One aim is to maximize the modal overlap between the two beams at a certain height such that a flat mirror can be used to couple a beam from one device into the other as in Fig.~\ref{concept}(c). A detailed analysis follows in Section~4.

\begin{figure*}[!t]
\begin{center}
\includegraphics[width=0.7\textwidth]{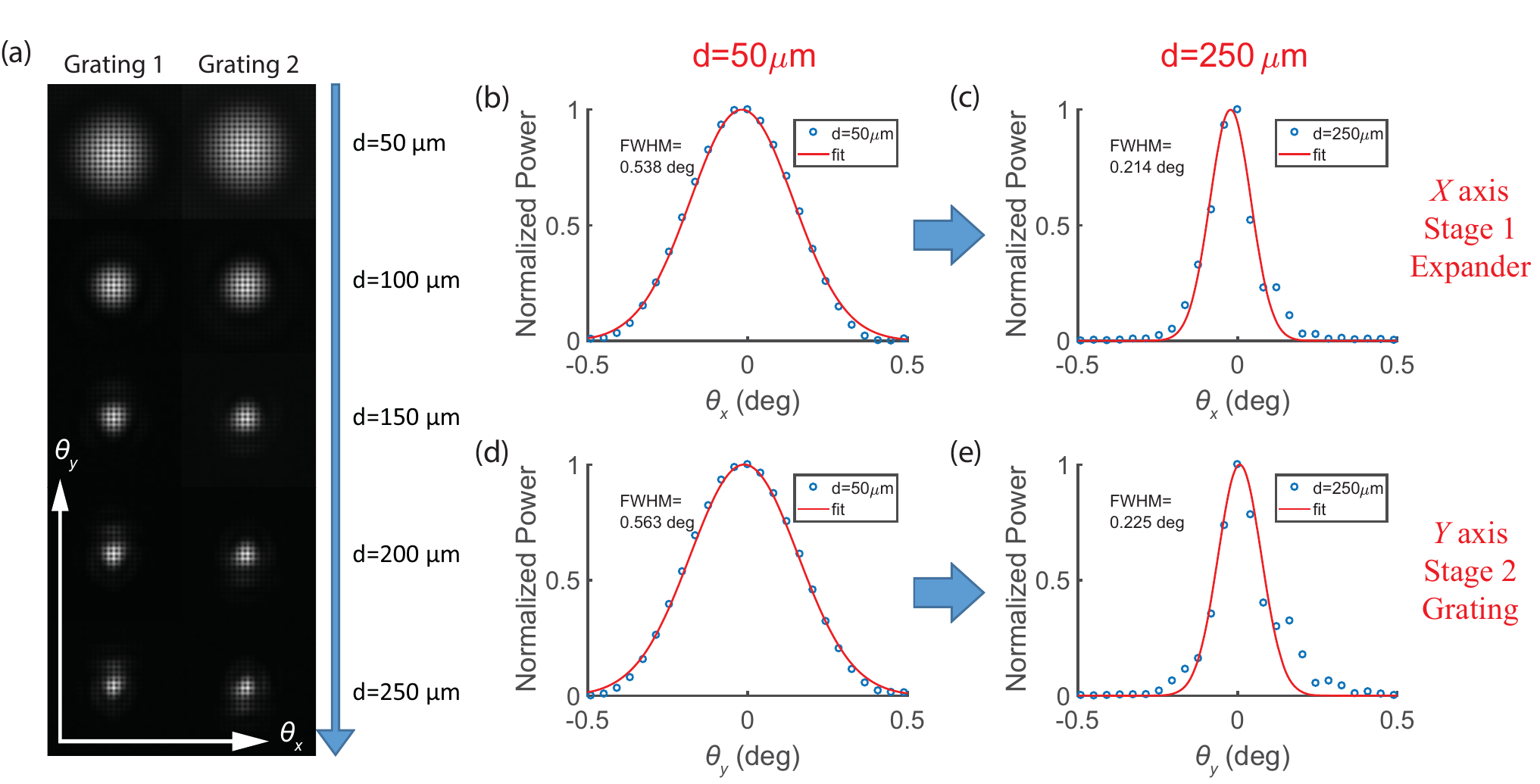}
\end{center}
\caption{Near-diffraction-limited Gaussian beam. (a) Zoomed-in back-focal-plane (BFP) images from mode converters 1 and 2 for different aperture diameters from 50~$\mu$m to 250~$\mu$m. Clipping the beam by the aperture increases diffraction. (b-e) Normalized powers of the BFP images of (a) at the center of each axis: (b,c) along the $\theta_x$ and (d,e) along the $\theta_y$ (blue circles: data, red lines: fitting Gaussian curves). (b) and (d) are cases of the aperture diameter $d\approx50~\mu$m, while (c) and (e) are cases of $d\approx250~\mu$m.
The uncertainties in the characterized FWHM is about $\pm0.005$$^{\circ}$ as determined by the Gaussian fit. }
\label{wavefront2}
\end{figure*}
Figure~\ref{wavefront2}(a) shows the zoomed-in view of the beam spots (rot0), from top to bottom, with different image-plane aperture diameters of $d\approx50~\mu$m to $250~\mu$m (left: grating~1, right: grating~2). Notice that as the aperture size increases in real space, the spot size in the Fourier domain reduces accordingly. Figures~\ref{wavefront2}(b)-\ref{wavefront2}(e) are the normalized powers cross-sectioned through the center of the spots (blue dots: data, red lines: fitting curves). Figures~\ref{wavefront2}(b) and \ref{wavefront2}(c) are along the $\theta_x$ for $d\approx50~\mu$m and $d\approx250~\mu$m, respectively, and Figs~\ref{wavefront2}(d) and \ref{wavefront2}(e) are along the $\theta_y$ for $d\approx50~\mu$m and $d\approx250~\mu$m, respectively. Note that the Gaussian distribution in the $x$ direction (or $\theta_x$) relates to the mode expansion from the waveguide to a slab mode (stage~1), while the Gaussian distribution in the $y$ direction (or $\theta_y$) is formed by the apodized grating during stage~2; they are formed by the two independent stages.
More importantly, for the large aperture of $d\approx250~\mu$m, as in Figs~\ref{wavefront2}(c) and \ref{wavefront2}(e), the measured angular FWHMs are close to the expected FWHM of a diffraction-limited beam, $i.e.$, FWHM$=\frac{2\sqrt{2\ln2}\lambda_0}{w_0\pi}=0.2094$$^{\circ}$ for $w_0=160~\mu$m and indicate a good beam collimation.

\section{Experiment 2: Out-coupling angles, modal overlap, and grating-to-grating coupling}
\begin{figure*}[!t]
\begin{center}
\includegraphics[width=0.9\textwidth]{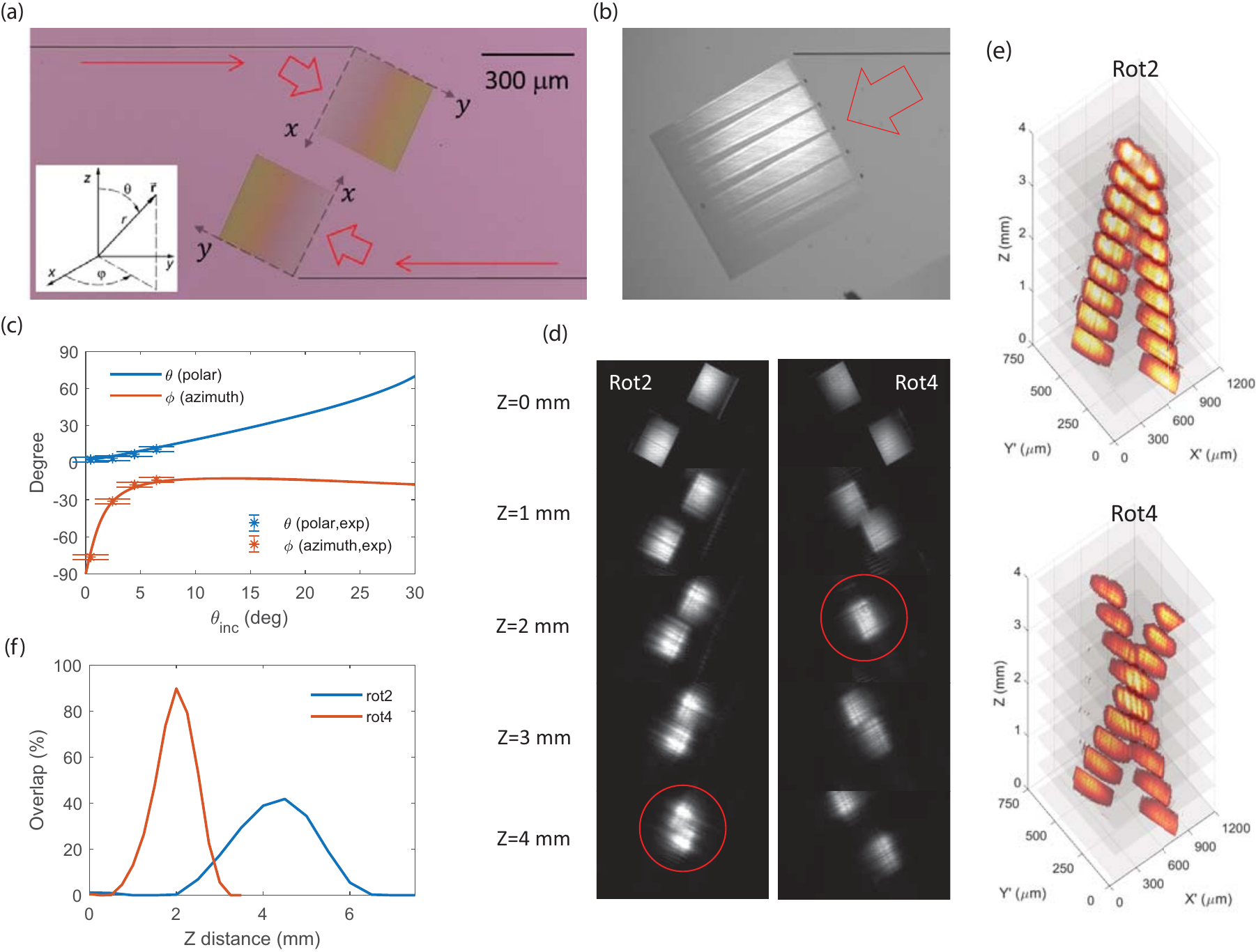}
\end{center}
\caption{Out-coupling angles and modal overlap.
(a) Microscope image of the two extreme mode converters. Inset shows the definitions of the polar ($\theta$) and azimuthal ($\phi$) angles.
(b) Zoomed-in converter image with scatterers embedded into the slab.
(c) Polar ($\theta$, blue) and azimuthal ($\phi$, orange) angles of the out-coupled beam as a function of incident angle $\theta_{\rm inc}$. The solid lines are analytical calculations and the points with error-bars are the measured angles.
Error bars are one standard deviation uncertainties propagated from beam center estimates in images.
(d) Microscope images of the two out-coupling beams at different distances $z$ (left: rot2, right: rot4). The red circles indicate the $z$-position of the maximum overlap.
(e) 3D stack images of (d).
(f) Measured modal overlap percentages of the two out-coupling beams as a function of $z$ (blue: rot2 and orange: rot4).}
\label{angles}
\end{figure*}
To integrate and optically couple the photonic chips with other systems in free-space, we can use the extreme mode converter as a building block establishing optical coupling between the systems. To explore such an engineering opportunity, we have placed two couplers with opposite orientations on the same photonic chip and arranged their  locations and out-coupling angles (polar and azimuthal) to create large beam overlap several millimeters above the chip. Finally, we have introduced a flat mirror at that location and conducted  grating-to-grating coupling experiments quantifying the coupling of the generated beam back to the photonic system.
Figure~\ref{angles}(a) shows the microscope image of the two extreme mode converters facing in the opposite directions with a center-to-center separation distance of approximately 475~$\mu$m.
Note that, in Fig.~\ref{concept}(b), the incident angle $\theta_{\rm inc}$ is defined for the angle between the slab mode $\beta_{\rm slab}$ and the grating vector $\Lambda$ ($y$ axis). The devices rot0, rot2, rot3, and rot4 have the designed $\theta_{\rm inc}=0^{\circ}, 2^{\circ}, 3^{\circ},$ and 4$^{\circ}$, respectively.
The tilt angle $\theta_{\rm tilt}$ is fixed at the nominal 28.18$^{\circ}$, and we rotate the gratings to adjust the $\theta_{\rm inc}$.
Also, note that the $xy$-axes are referenced to the grating, not in a global frame.
The azimuthal angle $\phi$ and the polar angle $\theta$ are defined with respect to the grating lines ($x$ axis) and the surface normal ($z$ axis), respectively (Inset of Fig.~\ref{angles}(a)).
To independently verify the propagation direction of the slab mode ($\theta_{\rm tilt}$) after the stage~1 (expander), we co-fabricated a nominally identical device in which we deliberately introduced a series of holes in front of the grating, serving as scatterers for the slab mode. Figure~\ref{angles}(b) shows the microscope image of such a device (rot0) with TE$_0$ input. The scatterers form shadows in the slab mode along its propagation direction, made visible by the grating. As expected, the angles of the shadows are close to $\theta_{\rm inc}\approx 0$$^{\circ}$ and the long, uniform-contrast shadows qualitatively indicate good collimation of the slab mode. This further validates the performance of the mode expander.

The polar $\theta$ and azimuthal $\phi$ angles of the  out-coupling beam can be engineered by varying the incident angle $\theta_{\rm inc}$, via grating rotation. Due to the momentum conservation in the grating plane, the out-coupling beam angles should follow
\begin{subequations}
\begin{align}
k\sin(\theta)\sin(\phi)&=\beta_y - \frac{2\pi m}{\Lambda} \quad (m: {\rm integer}),\\
k\sin(\theta)\cos(\phi)&=\beta_x.
\end{align}
\label{eq6}
\end{subequations}
where $k=2\pi/\lambda$, $\beta_y=\beta_{\rm slab}\cos(\theta_{\rm inc})$ and $\beta_x=\beta_{\rm slab}\sin(\theta_{\rm inc})$. The $\Lambda$ is the effective grating period and the propagation constant of the slab mode can be represented as $\beta_{\rm slab}=\frac{2\pi}{\lambda_0}n_{\rm eff}$. Rewriting Eq.~\ref{eq6} for the $\theta$ and $\phi$, we have
\begin{equation}
\sin{\theta}=\sqrt{
\left( n_{\rm eff}\sin\theta_{\rm inc} \right)^2 +
\left( n_{\rm eff}\cos\theta_{\rm inc} - \frac{m\lambda_0}{\Lambda} \right)^2},
\label{eq7}
\end{equation}
\begin{equation}
\tan{\phi}=\frac{\cos\theta_{\rm inc}-\frac{m\lambda_0}{n_{\rm eff}\Lambda}}
{\sin\theta_{\rm inc}}.
\label{eq8}
\end{equation}
To avoid losing light into multiple diffractions orders, and to create a near-vertical Gaussian beam for $\theta_{\rm inc}=0$, we have chosen $m=1$ and $1\gg n_{\rm eff} (\frac{\lambda_0}{\Lambda n_{\rm eff}}-1)>0$.
The blue and orange lines in Fig.~\ref{angles}(c) show the polar ($\theta$) and azimuthal ($\phi$) angles as a function of $\theta_{\rm inc}$, following Eq.~\ref{eq7} and Eq.~\ref{eq8}, respectively. For the parameters, $n_{\rm eff}=1.79$ and $\Lambda=425$~nm are assumed. The points represent the experimentally measured angles for $\theta_{\rm inc}=0$$^{\circ}$, 2$^{\circ}$, 4$^{\circ}$, and 6$^{\circ}$ and match with the analytical estimations.
Figure~\ref{angles}(d) shows the captured images of the rot2 (left) and rot4 (right) devices from this set at different $z$ positions. For each device, the $z$ positions of the maximum modal overlap are marked with red circles (around 4~mm for rot2 and 2~mm for the rot6). Figure~\ref{angles}(e) is the 3D stack images of Fig.~\ref{angles}(d) and shows the different azimuthal angles and mode overlap positions.

We have also experimentally estimated the overlap percentage of the two beams from each converter; we captured each of the beam images at different heights and used the following equation to extract the modal overlap:
\begin{equation}
{\rm Overlap}~(z)= \frac{\int\sqrt{I_1(z)I_2(z)}~dA}{\sqrt{\int I_1(z)~dA \int I_2(z)~dA}} \times100~\%
\label{eq9}
\end{equation}
The $I_1(z)$ and $I_2(z)$ are the beam intensity images from each device, separately.
The intensity measurement is not phase sensitive, and this equation provides an upper bound on the expected mode coupling loss by assuming perfect phase matching between the two beams, such as perfect collimation with beam waists at the mirror location.
The blue line in Fig.~\ref{angles}(f) shows the measured overlap percentage of rot2, and the orange line in Fig.~\ref{angles}(f) is that of rot4. As we can expect from Fig.~\ref{angles}(f), the $z$ position of the maximum overlap for rot2 and rot4 are $\approx4.5$~mm and $\approx2$~mm, respectively. We note that the maximum overlap percentage for rot4 is over 90~$\%$, fulfilling a necessary condition for low-loss grating-to-grating coupling, in which the radiating beam produced by one mode converter is reflected back to the chip (and a second converter) by placing a mirror at this position.

\begin{figure}[!t]
\includegraphics[width=0.5\textwidth]{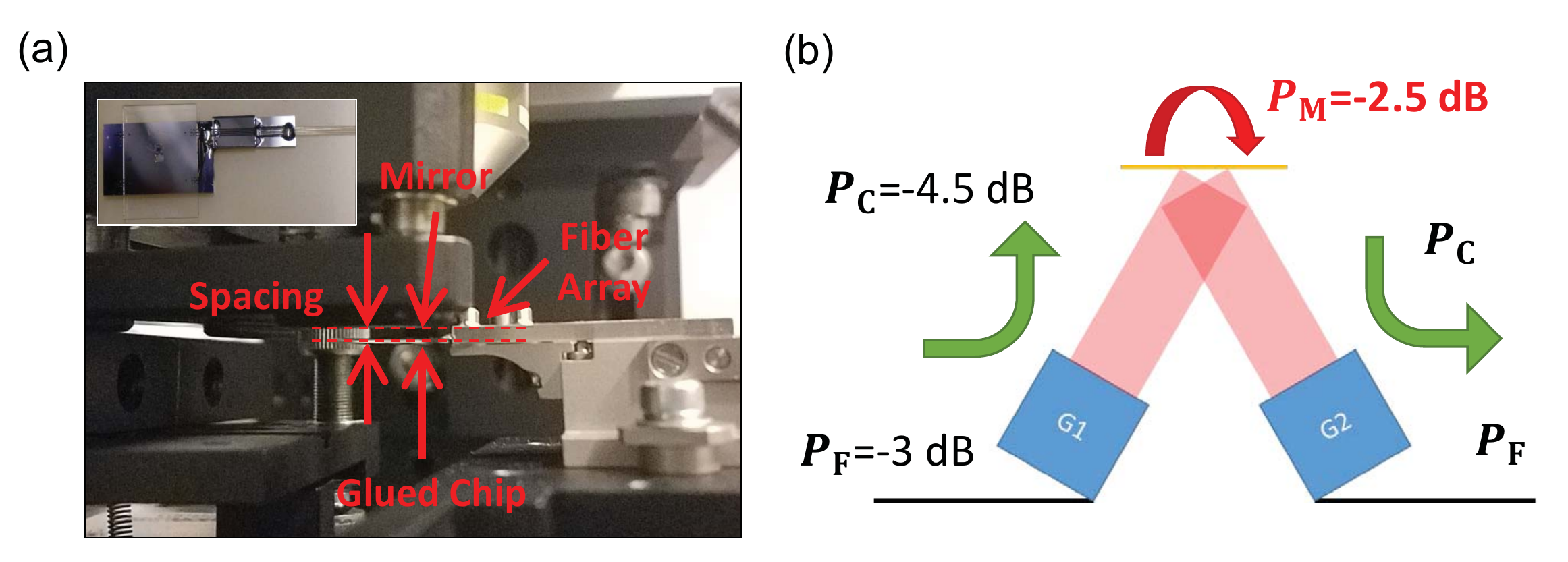}
\caption{Grating-to-grating coupling. (a) Measurement setup of the grating-to-grating coupling. The inset shows the top view of the chip, which is glued with a fiber array. (b) Schematic of loss components from each section.}
\label{mirror}
\end{figure}
Figure~\ref{mirror}(a) shows the experimental setup for the grating-to-grating coupling. The photonic chip is glued to a fiber array for input/output coupling, and a mirror is placed $\approx2$~mm above the chip, resulting in a maximum overlap for this rot4 device. The inset image shows the top view of the chip with the glued fiber array.
Figure~\ref{mirror}(b) shows the schematic of the entire device structure with key loss parameters; $P_{\rm F}$: fiber-to-chip edge-coupling loss, $P_{\rm C}$: extreme mode converter loss, and $P_{\rm M}$: mode-mismatch loss between the two beams.
The typical loss for the edge-coupling is $P_{\rm F}=-3~{\rm dB}\pm0.5$~dB, determined by multiple measurements using short waveguide loop-back structures connected to the inverse-tapered couplers. Here and below, the uncertainties are measured and propagated one standard deviation statistical uncertainties. Measuring the outgoing Gaussian beam power and assuming the above-mentioned value for the edge-coupling, the loss for the extreme mode converter is $P_{\rm C}=-4.5~{\rm dB}\pm0.5$~dB, corresponding to the overall loss of about $P_{\rm F}+P_{\rm C}=-7.5~{\rm dB}\pm0.5$~dB for each device. Using the setup in Fig.~\ref{mirror}(a), we could successfully couple the light back to the chip. By subtracting the independently-measured fiber-to-Gaussian-beam losses for each of the two mode converters from the total fiber-to-fiber loss, we have determined the excess loss due to the mode-mismatch between the two beams to be $P_{\rm M}=-2.5~{\rm dB}\pm1.0$~dB.

\section{Conclusion}
We have designed, numerically optimized, and experimentally demonstrated an extreme mode converter that can efficiently couple a few hundred-nanometer wide photonic waveguide mode to a hundred-micrometer wide free-space Gaussian beam. This expansion corresponds to an increase in mode area by a factor of $0.34 \times 10^6$. We have presented general guidelines for designing such a mode converter, and this approach can be applied to other types of mode conversions as well. Specifically, the evanescent expander offers a novel, optically-broadband approach for coupling single mode waveguides to wide slab modes with arbitrary profiles. We have successfully demonstrated the mode conversion experimentally and generated a Gaussian beam with a beam waist of $w_0\approx160~\mu$m. The converted Gaussian beam is well-collimated, approaching the diffraction limit, as confirmed by the BFP measurements. Furthermore, we have presented the ability to engineer the out-coupling beam direction and demonstrated a low loss grating-to-grating coupling using two converters on the same chip, with $\approx4~$mm of free space propagation distance. Our extreme mode converter can be used as a building block for the interaction of the photonic chip with other systems, enabling novel applications in atomic physics, biological and/or chemical sensing, LIDAR, biomedical health-care systems, and others.

\section*{Acknowledgments}
Dr. Sangsik Kim, Dr. Qing Li, and Dr. Alexander Yulaev acknowledge support under the Cooperative Research Agreement between the University of Maryland and the National Institute of Standards and Technology Center for Nanoscale Science and Technology, Award 70NANB10H193, through the University of Maryland.

\bibliography{noc_arXiv}

\begin{thebibliography}{29}%
\makeatletter
\providecommand \@ifxundefined [1]{%
 \@ifx{#1\undefined}
}%
\providecommand \@ifnum [1]{%
 \ifnum #1\expandafter \@firstoftwo
 \else \expandafter \@secondoftwo
 \fi
}%
\providecommand \@ifx [1]{%
 \ifx #1\expandafter \@firstoftwo
 \else \expandafter \@secondoftwo
 \fi
}%
\providecommand \natexlab [1]{#1}%
\providecommand \enquote  [1]{``#1''}%
\providecommand \bibnamefont  [1]{#1}%
\providecommand \bibfnamefont [1]{#1}%
\providecommand \citenamefont [1]{#1}%
\providecommand \href@noop [0]{\@secondoftwo}%
\providecommand \href [0]{\begingroup \@sanitize@url \@href}%
\providecommand \@href[1]{\@@startlink{#1}\@@href}%
\providecommand \@@href[1]{\endgroup#1\@@endlink}%
\providecommand \@sanitize@url [0]{\catcode `\\12\catcode `\$12\catcode
  `\&12\catcode `\#12\catcode `\^12\catcode `\_12\catcode `\%12\relax}%
\providecommand \@@startlink[1]{}%
\providecommand \@@endlink[0]{}%
\providecommand \url  [0]{\begingroup\@sanitize@url \@url }%
\providecommand \@url [1]{\endgroup\@href {#1}{\urlprefix }}%
\providecommand \urlprefix  [0]{URL }%
\providecommand \Eprint [0]{\href }%
\providecommand \doibase [0]{http://dx.doi.org/}%
\providecommand \selectlanguage [0]{\@gobble}%
\providecommand \bibinfo  [0]{\@secondoftwo}%
\providecommand \bibfield  [0]{\@secondoftwo}%
\providecommand \translation [1]{[#1]}%
\providecommand \BibitemOpen [0]{}%
\providecommand \bibitemStop [0]{}%
\providecommand \bibitemNoStop [0]{.\EOS\space}%
\providecommand \EOS [0]{\spacefactor3000\relax}%
\providecommand \BibitemShut  [1]{\csname bibitem#1\endcsname}%
\let\auto@bib@innerbib\@empty
\bibitem [{\citenamefont {Kitching}\ \emph {et~al.}(2016)\citenamefont
  {Kitching}, \citenamefont {Donley}, \citenamefont {Knappe}, \citenamefont
  {Hummon}, \citenamefont {Dellis}, \citenamefont {Sherman}, \citenamefont
  {Srinivasan}, \citenamefont {Aksyuk}, \citenamefont {Li}, \citenamefont
  {Westly}, \citenamefont {Roxworthy},\ and\ \citenamefont
  {Lal}}]{kitching2016nist}%
  \BibitemOpen
  \bibfield  {author} {\bibinfo {author} {\bibfnamefont {J.}~\bibnamefont
  {Kitching}}, \bibinfo {author} {\bibfnamefont {E.~A.}\ \bibnamefont
  {Donley}}, \bibinfo {author} {\bibfnamefont {S.}~\bibnamefont {Knappe}},
  \bibinfo {author} {\bibfnamefont {M.}~\bibnamefont {Hummon}}, \bibinfo
  {author} {\bibfnamefont {A.}~\bibnamefont {Dellis}}, \bibinfo {author}
  {\bibfnamefont {J.}~\bibnamefont {Sherman}}, \bibinfo {author} {\bibfnamefont
  {K.}~\bibnamefont {Srinivasan}}, \bibinfo {author} {\bibfnamefont {V.~A.}\
  \bibnamefont {Aksyuk}}, \bibinfo {author} {\bibfnamefont {Q.}~\bibnamefont
  {Li}}, \bibinfo {author} {\bibfnamefont {D.}~\bibnamefont {Westly}}, \bibinfo
  {author} {\bibfnamefont {B.}~\bibnamefont {Roxworthy}}, \ and\ \bibinfo
  {author} {\bibfnamefont {A.}~\bibnamefont {Lal}},\ }\href@noop {} {\bibfield
  {journal} {\bibinfo  {journal} {J. Phys. Conf. Ser.}\ }\textbf {\bibinfo
  {volume} {723}},\ \bibinfo {pages} {012056} (\bibinfo {year}
  {2016})}\BibitemShut {NoStop}%
\bibitem [{\citenamefont {Hummon}\ \emph {et~al.}()\citenamefont {Hummon},
  \citenamefont {Kang}, \citenamefont {Bopp}, \citenamefont {Li}, \citenamefont
  {Westly}, \citenamefont {Kim}, \citenamefont {Fredrick}, \citenamefont
  {Diddmas}, \citenamefont {Srinivasan}, \citenamefont {Aksyuk},\ and\
  \citenamefont {Kitching}}]{hummon2018photonic}%
  \BibitemOpen
  \bibfield  {author} {\bibinfo {author} {\bibfnamefont {M.}~\bibnamefont
  {Hummon}}, \bibinfo {author} {\bibfnamefont {S.}~\bibnamefont {Kang}},
  \bibinfo {author} {\bibfnamefont {D.}~\bibnamefont {Bopp}}, \bibinfo {author}
  {\bibfnamefont {Q.}~\bibnamefont {Li}}, \bibinfo {author} {\bibfnamefont
  {D.}~\bibnamefont {Westly}}, \bibinfo {author} {\bibfnamefont
  {S.}~\bibnamefont {Kim}}, \bibinfo {author} {\bibfnamefont {C.}~\bibnamefont
  {Fredrick}}, \bibinfo {author} {\bibfnamefont {S.}~\bibnamefont {Diddmas}},
  \bibinfo {author} {\bibfnamefont {K.}~\bibnamefont {Srinivasan}}, \bibinfo
  {author} {\bibfnamefont {V.}~\bibnamefont {Aksyuk}}, \ and\ \bibinfo {author}
  {\bibfnamefont {J.}~\bibnamefont {Kitching}},\ }\href@noop {} {\bibinfo
  {journal} {(under review)}\ }\BibitemShut {NoStop}%
\bibitem [{\citenamefont {Mehta}\ \emph {et~al.}(2016)\citenamefont {Mehta},
  \citenamefont {Bruzewicz}, \citenamefont {McConnell}, \citenamefont {Ram},
  \citenamefont {Sage},\ and\ \citenamefont
  {Chiaverini}}]{mehta2016integrated}%
  \BibitemOpen
\bibfield  {journal} {  }\bibfield  {author} {\bibinfo {author} {\bibfnamefont
  {K.~K.}\ \bibnamefont {Mehta}}, \bibinfo {author} {\bibfnamefont {C.~D.}\
  \bibnamefont {Bruzewicz}}, \bibinfo {author} {\bibfnamefont {R.}~\bibnamefont
  {McConnell}}, \bibinfo {author} {\bibfnamefont {R.~J.}\ \bibnamefont {Ram}},
  \bibinfo {author} {\bibfnamefont {J.~M.}\ \bibnamefont {Sage}}, \ and\
  \bibinfo {author} {\bibfnamefont {J.}~\bibnamefont {Chiaverini}},\
  }\href@noop {} {\bibfield  {journal} {\bibinfo  {journal} {Nat. Nanotech.}\
  }\textbf {\bibinfo {volume} {11}},\ \bibinfo {pages} {1066} (\bibinfo {year}
  {2016})}\BibitemShut {NoStop}%
\bibitem [{\citenamefont {Kohnen}\ \emph {et~al.}(2011)\citenamefont {Kohnen},
  \citenamefont {Succo}, \citenamefont {Petrov}, \citenamefont {Nyman},
  \citenamefont {Trupke},\ and\ \citenamefont {Hinds}}]{kohnen2011array}%
  \BibitemOpen
  \bibfield  {author} {\bibinfo {author} {\bibfnamefont {M.}~\bibnamefont
  {Kohnen}}, \bibinfo {author} {\bibfnamefont {M.}~\bibnamefont {Succo}},
  \bibinfo {author} {\bibfnamefont {P.}~\bibnamefont {Petrov}}, \bibinfo
  {author} {\bibfnamefont {R.}~\bibnamefont {Nyman}}, \bibinfo {author}
  {\bibfnamefont {M.}~\bibnamefont {Trupke}}, \ and\ \bibinfo {author}
  {\bibfnamefont {E.}~\bibnamefont {Hinds}},\ }\href@noop {} {\bibfield
  {journal} {\bibinfo  {journal} {Nat. Photon.}\ }\textbf {\bibinfo {volume}
  {5}},\ \bibinfo {pages} {35} (\bibinfo {year} {2011})}\BibitemShut {NoStop}%
\bibitem [{\citenamefont {Kippenberg}\ \emph {et~al.}(2011)\citenamefont
  {Kippenberg}, \citenamefont {Holzwarth},\ and\ \citenamefont
  {Diddams}}]{kippenberg2011microresonator}%
  \BibitemOpen
  \bibfield  {author} {\bibinfo {author} {\bibfnamefont {T.~J.}\ \bibnamefont
  {Kippenberg}}, \bibinfo {author} {\bibfnamefont {R.}~\bibnamefont
  {Holzwarth}}, \ and\ \bibinfo {author} {\bibfnamefont {S.}~\bibnamefont
  {Diddams}},\ }\href@noop {} {\bibfield  {journal} {\bibinfo  {journal}
  {Science}\ }\textbf {\bibinfo {volume} {332}},\ \bibinfo {pages} {555}
  (\bibinfo {year} {2011})}\BibitemShut {NoStop}%
\bibitem [{\citenamefont {Spencer}\ \emph {et~al.}(2017)\citenamefont
  {Spencer}, \citenamefont {Drake}, \citenamefont {Briles}, \citenamefont
  {Stone}, \citenamefont {Sinclair}, \citenamefont {Fredrick}, \citenamefont
  {Li}, \citenamefont {Westly}, \citenamefont {Ilic}, \citenamefont
  {Bluestone}, \citenamefont {Volet}, \citenamefont {Komljenovic},
  \citenamefont {Chang}, \citenamefont {Lee}, \citenamefont {Oh}, \citenamefont
  {Suh}, \citenamefont {Yang}, \citenamefont {Pfeiffer}, \citenamefont
  {Kippenberg}, \citenamefont {Norberg}, \citenamefont {Theogarajan},
  \citenamefont {Vahala}, \citenamefont {Newbury}, \citenamefont {Srinivasan},
  \citenamefont {Bowers}, \citenamefont {Diddams},\ and\ \citenamefont
  {Papp}}]{spencer2017integrated}%
  \BibitemOpen
  \bibfield  {author} {\bibinfo {author} {\bibfnamefont {D.~T.}\ \bibnamefont
  {Spencer}}, \bibinfo {author} {\bibfnamefont {T.}~\bibnamefont {Drake}},
  \bibinfo {author} {\bibfnamefont {T.~C.}\ \bibnamefont {Briles}}, \bibinfo
  {author} {\bibfnamefont {J.}~\bibnamefont {Stone}}, \bibinfo {author}
  {\bibfnamefont {L.~C.}\ \bibnamefont {Sinclair}}, \bibinfo {author}
  {\bibfnamefont {C.}~\bibnamefont {Fredrick}}, \bibinfo {author}
  {\bibfnamefont {Q.}~\bibnamefont {Li}}, \bibinfo {author} {\bibfnamefont
  {D.}~\bibnamefont {Westly}}, \bibinfo {author} {\bibfnamefont {B.~R.}\
  \bibnamefont {Ilic}}, \bibinfo {author} {\bibfnamefont {A.}~\bibnamefont
  {Bluestone}}, \bibinfo {author} {\bibfnamefont {N.}~\bibnamefont {Volet}},
  \bibinfo {author} {\bibfnamefont {T.}~\bibnamefont {Komljenovic}}, \bibinfo
  {author} {\bibfnamefont {L.}~\bibnamefont {Chang}}, \bibinfo {author}
  {\bibfnamefont {S.~H.}\ \bibnamefont {Lee}}, \bibinfo {author} {\bibfnamefont
  {D.~Y.}\ \bibnamefont {Oh}}, \bibinfo {author} {\bibfnamefont {M.-G.}\
  \bibnamefont {Suh}}, \bibinfo {author} {\bibfnamefont {K.~Y.}\ \bibnamefont
  {Yang}}, \bibinfo {author} {\bibfnamefont {M.~H.~P.}\ \bibnamefont
  {Pfeiffer}}, \bibinfo {author} {\bibfnamefont {T.~J.}\ \bibnamefont
  {Kippenberg}}, \bibinfo {author} {\bibfnamefont {E.}~\bibnamefont {Norberg}},
  \bibinfo {author} {\bibfnamefont {L.}~\bibnamefont {Theogarajan}}, \bibinfo
  {author} {\bibfnamefont {K.}~\bibnamefont {Vahala}}, \bibinfo {author}
  {\bibfnamefont {N.~R.}\ \bibnamefont {Newbury}}, \bibinfo {author}
  {\bibfnamefont {K.}~\bibnamefont {Srinivasan}}, \bibinfo {author}
  {\bibfnamefont {J.~E.}\ \bibnamefont {Bowers}}, \bibinfo {author}
  {\bibfnamefont {S.~A.}\ \bibnamefont {Diddams}}, \ and\ \bibinfo {author}
  {\bibfnamefont {S.~B.}\ \bibnamefont {Papp}},\ }\href@noop {} {\bibfield
  {journal} {\bibinfo  {journal} {arXiv:1708.05228}\ } (\bibinfo {year}
  {2017})}\BibitemShut {NoStop}%
\bibitem [{\citenamefont {Li}\ \emph {et~al.}(2017)\citenamefont {Li},
  \citenamefont {Briles}, \citenamefont {Westly}, \citenamefont {Drake},
  \citenamefont {Stone}, \citenamefont {Ilic}, \citenamefont {Diddams},
  \citenamefont {Papp},\ and\ \citenamefont {Srinivasan}}]{li2017stably}%
  \BibitemOpen
  \bibfield  {author} {\bibinfo {author} {\bibfnamefont {Q.}~\bibnamefont
  {Li}}, \bibinfo {author} {\bibfnamefont {T.~C.}\ \bibnamefont {Briles}},
  \bibinfo {author} {\bibfnamefont {D.~A.}\ \bibnamefont {Westly}}, \bibinfo
  {author} {\bibfnamefont {T.~E.}\ \bibnamefont {Drake}}, \bibinfo {author}
  {\bibfnamefont {J.~R.}\ \bibnamefont {Stone}}, \bibinfo {author}
  {\bibfnamefont {B.~R.}\ \bibnamefont {Ilic}}, \bibinfo {author}
  {\bibfnamefont {S.~A.}\ \bibnamefont {Diddams}}, \bibinfo {author}
  {\bibfnamefont {S.~B.}\ \bibnamefont {Papp}}, \ and\ \bibinfo {author}
  {\bibfnamefont {K.}~\bibnamefont {Srinivasan}},\ }\href@noop {} {\bibfield
  {journal} {\bibinfo  {journal} {Optica}\ }\textbf {\bibinfo {volume} {4}},\
  \bibinfo {pages} {193} (\bibinfo {year} {2017})}\BibitemShut {NoStop}%
\bibitem [{\citenamefont {Kim}\ \emph {et~al.}(2017)\citenamefont {Kim},
  \citenamefont {Han}, \citenamefont {Wang}, \citenamefont
  {Jaramillo-Villegas}, \citenamefont {Xue}, \citenamefont {Bao}, \citenamefont
  {Xuan}, \citenamefont {Leaird}, \citenamefont {Weiner},\ and\ \citenamefont
  {Qi}}]{kim2017dispersion}%
  \BibitemOpen
  \bibfield  {author} {\bibinfo {author} {\bibfnamefont {S.}~\bibnamefont
  {Kim}}, \bibinfo {author} {\bibfnamefont {K.}~\bibnamefont {Han}}, \bibinfo
  {author} {\bibfnamefont {C.}~\bibnamefont {Wang}}, \bibinfo {author}
  {\bibfnamefont {J.~A.}\ \bibnamefont {Jaramillo-Villegas}}, \bibinfo {author}
  {\bibfnamefont {X.}~\bibnamefont {Xue}}, \bibinfo {author} {\bibfnamefont
  {C.}~\bibnamefont {Bao}}, \bibinfo {author} {\bibfnamefont {Y.}~\bibnamefont
  {Xuan}}, \bibinfo {author} {\bibfnamefont {D.~E.}\ \bibnamefont {Leaird}},
  \bibinfo {author} {\bibfnamefont {A.~M.}\ \bibnamefont {Weiner}}, \ and\
  \bibinfo {author} {\bibfnamefont {M.}~\bibnamefont {Qi}},\ }\href@noop {}
  {\bibfield  {journal} {\bibinfo  {journal} {Nat. Commun.}\ }\textbf {\bibinfo
  {volume} {8}} (\bibinfo {year} {2017})}\BibitemShut {NoStop}%
\bibitem [{\citenamefont {Liang}\ \emph {et~al.}(2013)\citenamefont {Liang},
  \citenamefont {Clarke}, \citenamefont {Patel}, \citenamefont {Loncar},\ and\
  \citenamefont {Quan}}]{liang2013scalable}%
  \BibitemOpen
  \bibfield  {author} {\bibinfo {author} {\bibfnamefont {F.}~\bibnamefont
  {Liang}}, \bibinfo {author} {\bibfnamefont {N.}~\bibnamefont {Clarke}},
  \bibinfo {author} {\bibfnamefont {P.}~\bibnamefont {Patel}}, \bibinfo
  {author} {\bibfnamefont {M.}~\bibnamefont {Loncar}}, \ and\ \bibinfo {author}
  {\bibfnamefont {Q.}~\bibnamefont {Quan}},\ }\href@noop {} {\bibfield
  {journal} {\bibinfo  {journal} {Opt. Express}\ }\textbf {\bibinfo {volume}
  {21}},\ \bibinfo {pages} {32306} (\bibinfo {year} {2013})}\BibitemShut
  {NoStop}%
\bibitem [{\citenamefont {Fan}\ and\ \citenamefont
  {White}(2011)}]{fan2011optofluidic}%
  \BibitemOpen
  \bibfield  {author} {\bibinfo {author} {\bibfnamefont {X.}~\bibnamefont
  {Fan}}\ and\ \bibinfo {author} {\bibfnamefont {I.~M.}\ \bibnamefont
  {White}},\ }\href@noop {} {\bibfield  {journal} {\bibinfo  {journal} {Nat.
  Photon.}\ }\textbf {\bibinfo {volume} {5}},\ \bibinfo {pages} {591} (\bibinfo
  {year} {2011})}\BibitemShut {NoStop}%
\bibitem [{\citenamefont {Xu}\ \emph {et~al.}(2008)\citenamefont {Xu},
  \citenamefont {Densmore}, \citenamefont {Del{\^a}ge}, \citenamefont
  {Waldron}, \citenamefont {McKinnon}, \citenamefont {Janz}, \citenamefont
  {Lapointe}, \citenamefont {Lopinski}, \citenamefont {Mischki}, \citenamefont
  {Post}, \citenamefont {Cheben},\ and\ \citenamefont {Schmid}}]{xu2008folded}%
  \BibitemOpen
  \bibfield  {author} {\bibinfo {author} {\bibfnamefont {D.-X.}\ \bibnamefont
  {Xu}}, \bibinfo {author} {\bibfnamefont {A.}~\bibnamefont {Densmore}},
  \bibinfo {author} {\bibfnamefont {A.}~\bibnamefont {Del{\^a}ge}}, \bibinfo
  {author} {\bibfnamefont {P.}~\bibnamefont {Waldron}}, \bibinfo {author}
  {\bibfnamefont {R.}~\bibnamefont {McKinnon}}, \bibinfo {author}
  {\bibfnamefont {S.}~\bibnamefont {Janz}}, \bibinfo {author} {\bibfnamefont
  {J.}~\bibnamefont {Lapointe}}, \bibinfo {author} {\bibfnamefont
  {G.}~\bibnamefont {Lopinski}}, \bibinfo {author} {\bibfnamefont
  {T.}~\bibnamefont {Mischki}}, \bibinfo {author} {\bibfnamefont
  {E.}~\bibnamefont {Post}}, \bibinfo {author} {\bibfnamefont {P.}~\bibnamefont
  {Cheben}}, \ and\ \bibinfo {author} {\bibfnamefont {J.~H.}\ \bibnamefont
  {Schmid}},\ }\href@noop {} {\bibfield  {journal} {\bibinfo  {journal} {Opt.
  Express}\ }\textbf {\bibinfo {volume} {16}},\ \bibinfo {pages} {15137}
  (\bibinfo {year} {2008})}\BibitemShut {NoStop}%
\bibitem [{\citenamefont {Jokerst}\ \emph {et~al.}(2009)\citenamefont
  {Jokerst}, \citenamefont {Royal}, \citenamefont {Palit}, \citenamefont
  {Luan}, \citenamefont {Dhar},\ and\ \citenamefont {Tyler}}]{jokerst2009chip}%
  \BibitemOpen
  \bibfield  {author} {\bibinfo {author} {\bibfnamefont {N.}~\bibnamefont
  {Jokerst}}, \bibinfo {author} {\bibfnamefont {M.}~\bibnamefont {Royal}},
  \bibinfo {author} {\bibfnamefont {S.}~\bibnamefont {Palit}}, \bibinfo
  {author} {\bibfnamefont {L.}~\bibnamefont {Luan}}, \bibinfo {author}
  {\bibfnamefont {S.}~\bibnamefont {Dhar}}, \ and\ \bibinfo {author}
  {\bibfnamefont {T.}~\bibnamefont {Tyler}},\ }\href@noop {} {\bibfield
  {journal} {\bibinfo  {journal} {J. Biophotonics}\ }\textbf {\bibinfo {volume}
  {2}},\ \bibinfo {pages} {212} (\bibinfo {year} {2009})}\BibitemShut {NoStop}%
\bibitem [{\citenamefont {Lin}\ and\ \citenamefont
  {Crozier}(2013)}]{lin2013trapping}%
  \BibitemOpen
  \bibfield  {author} {\bibinfo {author} {\bibfnamefont {S.}~\bibnamefont
  {Lin}}\ and\ \bibinfo {author} {\bibfnamefont {K.~B.}\ \bibnamefont
  {Crozier}},\ }\href@noop {} {\bibfield  {journal} {\bibinfo  {journal} {ACS
  Nano}\ }\textbf {\bibinfo {volume} {7}},\ \bibinfo {pages} {1725} (\bibinfo
  {year} {2013})}\BibitemShut {NoStop}%
\bibitem [{\citenamefont {Thomson}\ \emph {et~al.}(2016)\citenamefont
  {Thomson}, \citenamefont {Zilkie}, \citenamefont {Bowers}, \citenamefont
  {Komljenovic}, \citenamefont {Reed}, \citenamefont {Vivien}, \citenamefont
  {Marris-Morini}, \citenamefont {Cassan}, \citenamefont {Virot}, \citenamefont
  {F{\'e}d{\'e}li}, \citenamefont {Hartmann}, \citenamefont {Schmid},
  \citenamefont {Xu}, \citenamefont {Boeuf}, \citenamefont {O{\^'}Brien},
  \citenamefont {Mashanovich},\ and\ \citenamefont
  {Nedeljkovic}}]{thomson2016roadmap}%
  \BibitemOpen
  \bibfield  {author} {\bibinfo {author} {\bibfnamefont {D.}~\bibnamefont
  {Thomson}}, \bibinfo {author} {\bibfnamefont {A.}~\bibnamefont {Zilkie}},
  \bibinfo {author} {\bibfnamefont {J.~E.}\ \bibnamefont {Bowers}}, \bibinfo
  {author} {\bibfnamefont {T.}~\bibnamefont {Komljenovic}}, \bibinfo {author}
  {\bibfnamefont {G.~T.}\ \bibnamefont {Reed}}, \bibinfo {author}
  {\bibfnamefont {L.}~\bibnamefont {Vivien}}, \bibinfo {author} {\bibfnamefont
  {D.}~\bibnamefont {Marris-Morini}}, \bibinfo {author} {\bibfnamefont
  {E.}~\bibnamefont {Cassan}}, \bibinfo {author} {\bibfnamefont
  {L.}~\bibnamefont {Virot}}, \bibinfo {author} {\bibfnamefont {J.-M.}\
  \bibnamefont {F{\'e}d{\'e}li}}, \bibinfo {author} {\bibfnamefont {J.-M.}\
  \bibnamefont {Hartmann}}, \bibinfo {author} {\bibfnamefont {J.~H.}\
  \bibnamefont {Schmid}}, \bibinfo {author} {\bibfnamefont {D.-X.}\
  \bibnamefont {Xu}}, \bibinfo {author} {\bibfnamefont {F.}~\bibnamefont
  {Boeuf}}, \bibinfo {author} {\bibfnamefont {P.}~\bibnamefont {O{\^'}Brien}},
  \bibinfo {author} {\bibfnamefont {G.~Z.}\ \bibnamefont {Mashanovich}}, \ and\
  \bibinfo {author} {\bibfnamefont {M.}~\bibnamefont {Nedeljkovic}},\
  }\href@noop {} {\bibfield  {journal} {\bibinfo  {journal} {J. Opt.}\ }\textbf
  {\bibinfo {volume} {18}},\ \bibinfo {pages} {073003} (\bibinfo {year}
  {2016})}\BibitemShut {NoStop}%
\bibitem [{\citenamefont {Agrell}\ \emph {et~al.}(2016)\citenamefont {Agrell},
  \citenamefont {Karlsson}, \citenamefont {Chraplyvy}, \citenamefont
  {Richardson}, \citenamefont {Krummrich}, \citenamefont {Winzer},
  \citenamefont {Roberts}, \citenamefont {Fischer}, \citenamefont {Savory},
  \citenamefont {Eggleton}, \citenamefont {Secondini}, \citenamefont
  {Kschischang}, \citenamefont {Lord}, \citenamefont {Prat}, \citenamefont
  {Tomkos}, \citenamefont {Bowers}, \citenamefont {Srinivasan}, \citenamefont
  {Brandt-Pearce},\ and\ \citenamefont {Gisin}}]{agrell2016roadmap}%
  \BibitemOpen
  \bibfield  {author} {\bibinfo {author} {\bibfnamefont {E.}~\bibnamefont
  {Agrell}}, \bibinfo {author} {\bibfnamefont {M.}~\bibnamefont {Karlsson}},
  \bibinfo {author} {\bibfnamefont {A.~R.}\ \bibnamefont {Chraplyvy}}, \bibinfo
  {author} {\bibfnamefont {D.~J.}\ \bibnamefont {Richardson}}, \bibinfo
  {author} {\bibfnamefont {P.~M.}\ \bibnamefont {Krummrich}}, \bibinfo {author}
  {\bibfnamefont {P.}~\bibnamefont {Winzer}}, \bibinfo {author} {\bibfnamefont
  {K.}~\bibnamefont {Roberts}}, \bibinfo {author} {\bibfnamefont {J.~K.}\
  \bibnamefont {Fischer}}, \bibinfo {author} {\bibfnamefont {S.~J.}\
  \bibnamefont {Savory}}, \bibinfo {author} {\bibfnamefont {B.~J.}\
  \bibnamefont {Eggleton}}, \bibinfo {author} {\bibfnamefont {M.}~\bibnamefont
  {Secondini}}, \bibinfo {author} {\bibfnamefont {F.~R.}\ \bibnamefont
  {Kschischang}}, \bibinfo {author} {\bibfnamefont {A.}~\bibnamefont {Lord}},
  \bibinfo {author} {\bibfnamefont {J.}~\bibnamefont {Prat}}, \bibinfo {author}
  {\bibfnamefont {I.}~\bibnamefont {Tomkos}}, \bibinfo {author} {\bibfnamefont
  {J.~E.}\ \bibnamefont {Bowers}}, \bibinfo {author} {\bibfnamefont
  {S.}~\bibnamefont {Srinivasan}}, \bibinfo {author} {\bibfnamefont
  {M.}~\bibnamefont {Brandt-Pearce}}, \ and\ \bibinfo {author} {\bibfnamefont
  {N.}~\bibnamefont {Gisin}},\ }\href@noop {} {\bibfield  {journal} {\bibinfo
  {journal} {J. Opt}\ }\textbf {\bibinfo {volume} {18}},\ \bibinfo {pages}
  {063002} (\bibinfo {year} {2016})}\BibitemShut {NoStop}%
\bibitem [{\citenamefont {Jahani}\ \emph {et~al.}(2017)\citenamefont {Jahani},
  \citenamefont {Kim}, \citenamefont {Atkinson}, \citenamefont {Wirth},
  \citenamefont {Kalhor}, \citenamefont {Newman}, \citenamefont {Shekhar},
  \citenamefont {Han}, \citenamefont {Van}, \citenamefont {DeCorby},
  \citenamefont {Chrostowski}, \citenamefont {Qi},\ and\ \citenamefont
  {Jacob}}]{jahani2017photonic}%
  \BibitemOpen
  \bibfield  {author} {\bibinfo {author} {\bibfnamefont {S.}~\bibnamefont
  {Jahani}}, \bibinfo {author} {\bibfnamefont {S.}~\bibnamefont {Kim}},
  \bibinfo {author} {\bibfnamefont {J.}~\bibnamefont {Atkinson}}, \bibinfo
  {author} {\bibfnamefont {J.~C.}\ \bibnamefont {Wirth}}, \bibinfo {author}
  {\bibfnamefont {F.}~\bibnamefont {Kalhor}}, \bibinfo {author} {\bibfnamefont
  {W.~D.}\ \bibnamefont {Newman}}, \bibinfo {author} {\bibfnamefont
  {P.}~\bibnamefont {Shekhar}}, \bibinfo {author} {\bibfnamefont
  {K.}~\bibnamefont {Han}}, \bibinfo {author} {\bibfnamefont {V.}~\bibnamefont
  {Van}}, \bibinfo {author} {\bibfnamefont {R.~G.}\ \bibnamefont {DeCorby}},
  \bibinfo {author} {\bibfnamefont {L.}~\bibnamefont {Chrostowski}}, \bibinfo
  {author} {\bibfnamefont {M.}~\bibnamefont {Qi}}, \ and\ \bibinfo {author}
  {\bibfnamefont {Z.}~\bibnamefont {Jacob}},\ }\href@noop {} {\bibfield
  {journal} {\bibinfo  {journal} {arXiv:1701.03093}\ } (\bibinfo {year}
  {2017})}\BibitemShut {NoStop}%
\bibitem [{\citenamefont {Doylend}\ \emph {et~al.}(2011)\citenamefont
  {Doylend}, \citenamefont {Heck}, \citenamefont {Bovington}, \citenamefont
  {Peters}, \citenamefont {Coldren},\ and\ \citenamefont
  {Bowers}}]{doylend2011two}%
  \BibitemOpen
  \bibfield  {author} {\bibinfo {author} {\bibfnamefont {J.~K.}\ \bibnamefont
  {Doylend}}, \bibinfo {author} {\bibfnamefont {M.}~\bibnamefont {Heck}},
  \bibinfo {author} {\bibfnamefont {J.~T.}\ \bibnamefont {Bovington}}, \bibinfo
  {author} {\bibfnamefont {J.~D.}\ \bibnamefont {Peters}}, \bibinfo {author}
  {\bibfnamefont {L.}~\bibnamefont {Coldren}}, \ and\ \bibinfo {author}
  {\bibfnamefont {J.}~\bibnamefont {Bowers}},\ }\href@noop {} {\bibfield
  {journal} {\bibinfo  {journal} {Opt. Express}\ }\textbf {\bibinfo {volume}
  {19}},\ \bibinfo {pages} {21595} (\bibinfo {year} {2011})}\BibitemShut
  {NoStop}%
\bibitem [{\citenamefont {Sun}\ \emph {et~al.}(2013)\citenamefont {Sun},
  \citenamefont {Timurdogan}, \citenamefont {Yaacobi}, \citenamefont
  {Hosseini},\ and\ \citenamefont {Watts}}]{sun2013large}%
  \BibitemOpen
  \bibfield  {author} {\bibinfo {author} {\bibfnamefont {J.}~\bibnamefont
  {Sun}}, \bibinfo {author} {\bibfnamefont {E.}~\bibnamefont {Timurdogan}},
  \bibinfo {author} {\bibfnamefont {A.}~\bibnamefont {Yaacobi}}, \bibinfo
  {author} {\bibfnamefont {E.~S.}\ \bibnamefont {Hosseini}}, \ and\ \bibinfo
  {author} {\bibfnamefont {M.~R.}\ \bibnamefont {Watts}},\ }\href@noop {}
  {\bibfield  {journal} {\bibinfo  {journal} {Nature}\ }\textbf {\bibinfo
  {volume} {493}},\ \bibinfo {pages} {195} (\bibinfo {year}
  {2013})}\BibitemShut {NoStop}%
\bibitem [{\citenamefont {Poulton}\ \emph {et~al.}(2017)\citenamefont
  {Poulton}, \citenamefont {Byrd}, \citenamefont {Raval}, \citenamefont {Su},
  \citenamefont {Li}, \citenamefont {Timurdogan}, \citenamefont {Coolbaugh},
  \citenamefont {Vermeulen},\ and\ \citenamefont {Watts}}]{poulton2017large}%
  \BibitemOpen
  \bibfield  {author} {\bibinfo {author} {\bibfnamefont {C.~V.}\ \bibnamefont
  {Poulton}}, \bibinfo {author} {\bibfnamefont {M.~J.}\ \bibnamefont {Byrd}},
  \bibinfo {author} {\bibfnamefont {M.}~\bibnamefont {Raval}}, \bibinfo
  {author} {\bibfnamefont {Z.}~\bibnamefont {Su}}, \bibinfo {author}
  {\bibfnamefont {N.}~\bibnamefont {Li}}, \bibinfo {author} {\bibfnamefont
  {E.}~\bibnamefont {Timurdogan}}, \bibinfo {author} {\bibfnamefont
  {D.}~\bibnamefont {Coolbaugh}}, \bibinfo {author} {\bibfnamefont
  {D.}~\bibnamefont {Vermeulen}}, \ and\ \bibinfo {author} {\bibfnamefont
  {M.~R.}\ \bibnamefont {Watts}},\ }\href@noop {} {\bibfield  {journal}
  {\bibinfo  {journal} {Opt. Lett.}\ }\textbf {\bibinfo {volume} {42}},\
  \bibinfo {pages} {21} (\bibinfo {year} {2017})}\BibitemShut {NoStop}%
\bibitem [{\citenamefont {Mekis}\ \emph {et~al.}(2011)\citenamefont {Mekis},
  \citenamefont {Gloeckner}, \citenamefont {Masini}, \citenamefont {Narasimha},
  \citenamefont {Pinguet}, \citenamefont {Sahni},\ and\ \citenamefont
  {De~Dobbelaere}}]{mekis2011grating}%
  \BibitemOpen
  \bibfield  {author} {\bibinfo {author} {\bibfnamefont {A.}~\bibnamefont
  {Mekis}}, \bibinfo {author} {\bibfnamefont {S.}~\bibnamefont {Gloeckner}},
  \bibinfo {author} {\bibfnamefont {G.}~\bibnamefont {Masini}}, \bibinfo
  {author} {\bibfnamefont {A.}~\bibnamefont {Narasimha}}, \bibinfo {author}
  {\bibfnamefont {T.}~\bibnamefont {Pinguet}}, \bibinfo {author} {\bibfnamefont
  {S.}~\bibnamefont {Sahni}}, \ and\ \bibinfo {author} {\bibfnamefont
  {P.}~\bibnamefont {De~Dobbelaere}},\ }\href@noop {} {\bibfield  {journal}
  {\bibinfo  {journal} {IEEE J. Sel. Topics Quantum Electron.}\ }\textbf
  {\bibinfo {volume} {17}},\ \bibinfo {pages} {597} (\bibinfo {year}
  {2011})}\BibitemShut {NoStop}%
\bibitem [{\citenamefont {Chen}\ \emph {et~al.}(2010)\citenamefont {Chen},
  \citenamefont {Li}, \citenamefont {Fung}, \citenamefont {Lo},\ and\
  \citenamefont {Tsang}}]{chen2010apodized}%
  \BibitemOpen
  \bibfield  {author} {\bibinfo {author} {\bibfnamefont {X.}~\bibnamefont
  {Chen}}, \bibinfo {author} {\bibfnamefont {C.}~\bibnamefont {Li}}, \bibinfo
  {author} {\bibfnamefont {C.~K.}\ \bibnamefont {Fung}}, \bibinfo {author}
  {\bibfnamefont {S.~M.}\ \bibnamefont {Lo}}, \ and\ \bibinfo {author}
  {\bibfnamefont {H.~K.}\ \bibnamefont {Tsang}},\ }\href@noop {} {\bibfield
  {journal} {\bibinfo  {journal} {IEEE Photon. Technol. Lett}\ }\textbf
  {\bibinfo {volume} {22}},\ \bibinfo {pages} {1156} (\bibinfo {year}
  {2010})}\BibitemShut {NoStop}%
\bibitem [{\citenamefont {Ding}\ \emph {et~al.}(2013)\citenamefont {Ding},
  \citenamefont {Ou},\ and\ \citenamefont {Peucheret}}]{ding2013ultrahigh}%
  \BibitemOpen
  \bibfield  {author} {\bibinfo {author} {\bibfnamefont {Y.}~\bibnamefont
  {Ding}}, \bibinfo {author} {\bibfnamefont {H.}~\bibnamefont {Ou}}, \ and\
  \bibinfo {author} {\bibfnamefont {C.}~\bibnamefont {Peucheret}},\ }\href@noop
  {} {\bibfield  {journal} {\bibinfo  {journal} {Opt. Lett.}\ }\textbf
  {\bibinfo {volume} {38}},\ \bibinfo {pages} {2732} (\bibinfo {year}
  {2013})}\BibitemShut {NoStop}%
\bibitem [{\citenamefont {Mehta}\ and\ \citenamefont
  {Ram}(2017)}]{mehta2017precise}%
  \BibitemOpen
  \bibfield  {author} {\bibinfo {author} {\bibfnamefont {K.~K.}\ \bibnamefont
  {Mehta}}\ and\ \bibinfo {author} {\bibfnamefont {R.~J.}\ \bibnamefont
  {Ram}},\ }\href@noop {} {\bibfield  {journal} {\bibinfo  {journal} {Sci.
  Rep.}\ }\textbf {\bibinfo {volume} {7}},\ \bibinfo {pages} {2019} (\bibinfo
  {year} {2017})}\BibitemShut {NoStop}%
\bibitem [{\citenamefont {Xu}\ \emph {et~al.}(2012)\citenamefont {Xu},
  \citenamefont {Subbaraman}, \citenamefont {Covey}, \citenamefont {Kwong},
  \citenamefont {Hosseini},\ and\ \citenamefont {Chen}}]{xu2012complementary}%
  \BibitemOpen
  \bibfield  {author} {\bibinfo {author} {\bibfnamefont {X.}~\bibnamefont
  {Xu}}, \bibinfo {author} {\bibfnamefont {H.}~\bibnamefont {Subbaraman}},
  \bibinfo {author} {\bibfnamefont {J.}~\bibnamefont {Covey}}, \bibinfo
  {author} {\bibfnamefont {D.}~\bibnamefont {Kwong}}, \bibinfo {author}
  {\bibfnamefont {A.}~\bibnamefont {Hosseini}}, \ and\ \bibinfo {author}
  {\bibfnamefont {R.~T.}\ \bibnamefont {Chen}},\ }\href@noop {} {\bibfield
  {journal} {\bibinfo  {journal} {Appl. Phys. Lett.}\ }\textbf {\bibinfo
  {volume} {101}},\ \bibinfo {pages} {031109} (\bibinfo {year}
  {2012})}\BibitemShut {NoStop}%
\bibitem [{\citenamefont {Song}\ \emph {et~al.}(2015)\citenamefont {Song},
  \citenamefont {Doany}, \citenamefont {Medhin}, \citenamefont {Dupuis},
  \citenamefont {Lee},\ and\ \citenamefont {Libsch}}]{song2015polarization}%
  \BibitemOpen
  \bibfield  {author} {\bibinfo {author} {\bibfnamefont {J.~H.}\ \bibnamefont
  {Song}}, \bibinfo {author} {\bibfnamefont {F.~E.}\ \bibnamefont {Doany}},
  \bibinfo {author} {\bibfnamefont {A.~K.}\ \bibnamefont {Medhin}}, \bibinfo
  {author} {\bibfnamefont {N.}~\bibnamefont {Dupuis}}, \bibinfo {author}
  {\bibfnamefont {B.~G.}\ \bibnamefont {Lee}}, \ and\ \bibinfo {author}
  {\bibfnamefont {F.~R.}\ \bibnamefont {Libsch}},\ }\href@noop {} {\bibfield
  {journal} {\bibinfo  {journal} {Opt. Lett.}\ }\textbf {\bibinfo {volume}
  {40}},\ \bibinfo {pages} {3941} (\bibinfo {year} {2015})}\BibitemShut
  {NoStop}%
\bibitem [{\citenamefont {Vermeulen}\ \emph {et~al.}(2010)\citenamefont
  {Vermeulen}, \citenamefont {Selvaraja}, \citenamefont {Verheyen},
  \citenamefont {Lepage}, \citenamefont {Bogaerts}, \citenamefont {Absil},
  \citenamefont {Van~Thourhout},\ and\ \citenamefont
  {Roelkens}}]{vermeulen2010high}%
  \BibitemOpen
  \bibfield  {author} {\bibinfo {author} {\bibfnamefont {D.}~\bibnamefont
  {Vermeulen}}, \bibinfo {author} {\bibfnamefont {S.}~\bibnamefont
  {Selvaraja}}, \bibinfo {author} {\bibfnamefont {P.}~\bibnamefont {Verheyen}},
  \bibinfo {author} {\bibfnamefont {G.}~\bibnamefont {Lepage}}, \bibinfo
  {author} {\bibfnamefont {W.}~\bibnamefont {Bogaerts}}, \bibinfo {author}
  {\bibfnamefont {P.}~\bibnamefont {Absil}}, \bibinfo {author} {\bibfnamefont
  {D.}~\bibnamefont {Van~Thourhout}}, \ and\ \bibinfo {author} {\bibfnamefont
  {G.}~\bibnamefont {Roelkens}},\ }\href@noop {} {\bibfield  {journal}
  {\bibinfo  {journal} {Opt. Express}\ }\textbf {\bibinfo {volume} {18}},\
  \bibinfo {pages} {18278} (\bibinfo {year} {2010})}\BibitemShut {NoStop}%
\bibitem [{\citenamefont {Chen}\ \emph {et~al.}(2012)\citenamefont {Chen},
  \citenamefont {Xu}, \citenamefont {Cheng}, \citenamefont {Fung},\ and\
  \citenamefont {Tsang}}]{chen2012wideband}%
  \BibitemOpen
  \bibfield  {author} {\bibinfo {author} {\bibfnamefont {X.}~\bibnamefont
  {Chen}}, \bibinfo {author} {\bibfnamefont {K.}~\bibnamefont {Xu}}, \bibinfo
  {author} {\bibfnamefont {Z.}~\bibnamefont {Cheng}}, \bibinfo {author}
  {\bibfnamefont {C.~K.}\ \bibnamefont {Fung}}, \ and\ \bibinfo {author}
  {\bibfnamefont {H.~K.}\ \bibnamefont {Tsang}},\ }\href@noop {} {\bibfield
  {journal} {\bibinfo  {journal} {Opt. Lett.}\ }\textbf {\bibinfo {volume}
  {37}},\ \bibinfo {pages} {3483} (\bibinfo {year} {2012})}\BibitemShut
  {NoStop}%
\bibitem [{\citenamefont {Halir}\ \emph {et~al.}(2010)\citenamefont {Halir},
  \citenamefont {Cheben}, \citenamefont {Schmid}, \citenamefont {Ma},
  \citenamefont {Bedard}, \citenamefont {Janz}, \citenamefont {Xu},
  \citenamefont {Densmore}, \citenamefont {Lapointe},\ and\ \citenamefont
  {Molina-Fern{\'a}ndez}}]{halir2010continuously}%
  \BibitemOpen
  \bibfield  {author} {\bibinfo {author} {\bibfnamefont {R.}~\bibnamefont
  {Halir}}, \bibinfo {author} {\bibfnamefont {P.}~\bibnamefont {Cheben}},
  \bibinfo {author} {\bibfnamefont {J.}~\bibnamefont {Schmid}}, \bibinfo
  {author} {\bibfnamefont {R.}~\bibnamefont {Ma}}, \bibinfo {author}
  {\bibfnamefont {D.}~\bibnamefont {Bedard}}, \bibinfo {author} {\bibfnamefont
  {S.}~\bibnamefont {Janz}}, \bibinfo {author} {\bibfnamefont {D.-X.}\
  \bibnamefont {Xu}}, \bibinfo {author} {\bibfnamefont {A.}~\bibnamefont
  {Densmore}}, \bibinfo {author} {\bibfnamefont {J.}~\bibnamefont {Lapointe}},
  \ and\ \bibinfo {author} {\bibfnamefont {I.}~\bibnamefont
  {Molina-Fern{\'a}ndez}},\ }\href@noop {} {\bibfield  {journal} {\bibinfo
  {journal} {Opt. Lett.}\ }\textbf {\bibinfo {volume} {35}},\ \bibinfo {pages}
  {3243} (\bibinfo {year} {2010})}\BibitemShut {NoStop}%
\bibitem [{Com()}]{ComsolOpt}%
  \BibitemOpen
  \href@noop {} {}\bibinfo {howpublished} {COMSOL Multiphysics Acoustic
  Optimization Model,
  \url{https://www.comsol.com/model/optimizing-the-shape-of-a-horn-4353}}\BibitemShut
  {NoStop}%
\end{thebibliography}%

\end{document}